\documentclass[11pt]{article}%
\usepackage{graphicx}
\usepackage[margin=1in]{geometry}
\usepackage{authblk}
\usepackage{amsmath}
\usepackage{amsfonts}
\usepackage{amssymb}
\usepackage{natbib}
\usepackage{color}
\usepackage[normalem]{ulem}
\providecommand{\U}[1]{\protect\rule{.1in}{.1in}}
\providecommand{\U}[1]{\protect\rule{.1in}{.1in}}

\newtheorem{remark}{Remark}
\begin{document}

\title{The Generalized Carrier-Greenspan Transform for the shallow water system with arbitrary initial and boundary conditions}
\author{Alexei Rybkin$^1$
\and Dmitry Nicolsky$^2$
\and Efim Pelinovsky$^{3,4,5}$
\and Maxwell Buckel$^1$}
\date{%
    $^1$Department of Mathematics and Statistics, University of Alaska Fairbanks, USA\\%
    $^2$Geophysical Institute, University of Alaska Fairbanks, USA\\%
    $^3$National Research University - Higher School of Economics, Moscow, Russia\\%
    $^4$Institute of Applied Physics, Nizhny Novgorod, Russia\\%
    $^5$Nizhny Novgorod State Technical University n.a. R.E. Alekseev, Nizhny Novgorod, Russia\\%
    \today
}
\maketitle

\begin{abstract}
We put forward a solution to the initial boundary value (IBV) problem for the nonlinear shallow water system in inclined channels of arbitrary cross-section
by means of the generalized Carrier-Greenspan hodograph transform \citep{Rybkin}. Since the Carrier-Greenspan transform, while linearizing the shallow water system, seriously entangles the IBV in the hodograph plane, all previous solutions required some restrictive assumptions on the IBV conditions, e.g., zero initial velocity, smallness of boundary conditions. For arbitrary non-breaking initial conditions in the physical space, we present an explicit formula for equivalent IBV conditions in the hodograph plane, which can readily be treated by conventional methods. Our procedure, which we call the method of data projection, is based on the Taylor formula and allows us to reduce the transformed IBV data given on curves in the hodograph plane to the equivalent data on lines. Our method works equally well for any inclined bathymetry (not only plane beaches) and, moreover, is fully analytical for U-shaped bays. Numerical simulations show that our method is very robust and can be used to give express forecasting of tsunami wave inundation in narrow bays and fjords\footnote{To appear in Water Waves (2020)}.
\end{abstract}

\affil[1]{}
\affil[2]{}
\affil[3]{}
\affil[4]{}
\affil[5]{Special Research Bureau for Automation of Marine Researches, Yuzhno-Sakhalinsk, Russia}
\affil[6]{}

\section{Introduction}

\bigskip
Walter Craig made an outstanding contribution towards development of nonlinear theories of the long wave dynamics in fluids of the variable depths \citep{Craig94,Craig04,Craig05a,Craig05b,Craig06,Craig07}. Here we present a new solution of the nonlinear shallow-water equations for long waves, tsunamis, in the inclined channels of variable depth.
As a motion of viscous fluid, tsunami waves are described by the Navier-Stokes equations, a highly nonlinear 3+1 (three spatial and one temporal derivatives) system, which is notoriously hard to analyze even numerically. However, in many important cases some extra assumptions lead to considerable simplifications. For instance, assuming that depth/wavelength, wave height/depth are small and truncating the Taylor expansions of nonlinear terms produce a whole zoo of approximations commonly called shallow water equations (e.g. Korteweg--de Vries, Boussinesq, Saint--Venant, to name just three). Further assumptions that the vertical velocity is small and no vorticity effectively reduce the Navier--Stokes equations to the (2+1) shallow water-wave equations (SWE) which provide an accurate model for predicting tsunami wave behavior \citep{Synolakis91,Synolakis06,NTHMP12,Kanoglu15,TsunamiDynamics15}. Still, for general bathymetries this model allows us to analyze tsunami wave run-ups (our main concern) only numerically (for an analytical solution for a specific bathymetry see \citep{Synolakis08}). For a complete analysis of tsunami hydrodynamics, modeling, and forecasting, we refer the reader to \citet{Kanoglu15,Pelinovsky06,MadsenFuhrman}, and \citet{Synolakis06}. Mathematically rigorous treatment of SWE including the well-posedness and exact solutions can be found in \citet{Dobrokhotov10a,Dobrokhotov10b,Dobrokhotov13}, \citet{Alekseenko17} and references therein. It is worth mentioning that IVP for SWE have been treated in \citet{Chugunov14,Chugunov20} using a perturbation approach.

\begin{figure}[ptb]
\centering
\includegraphics[width=0.9\textwidth]{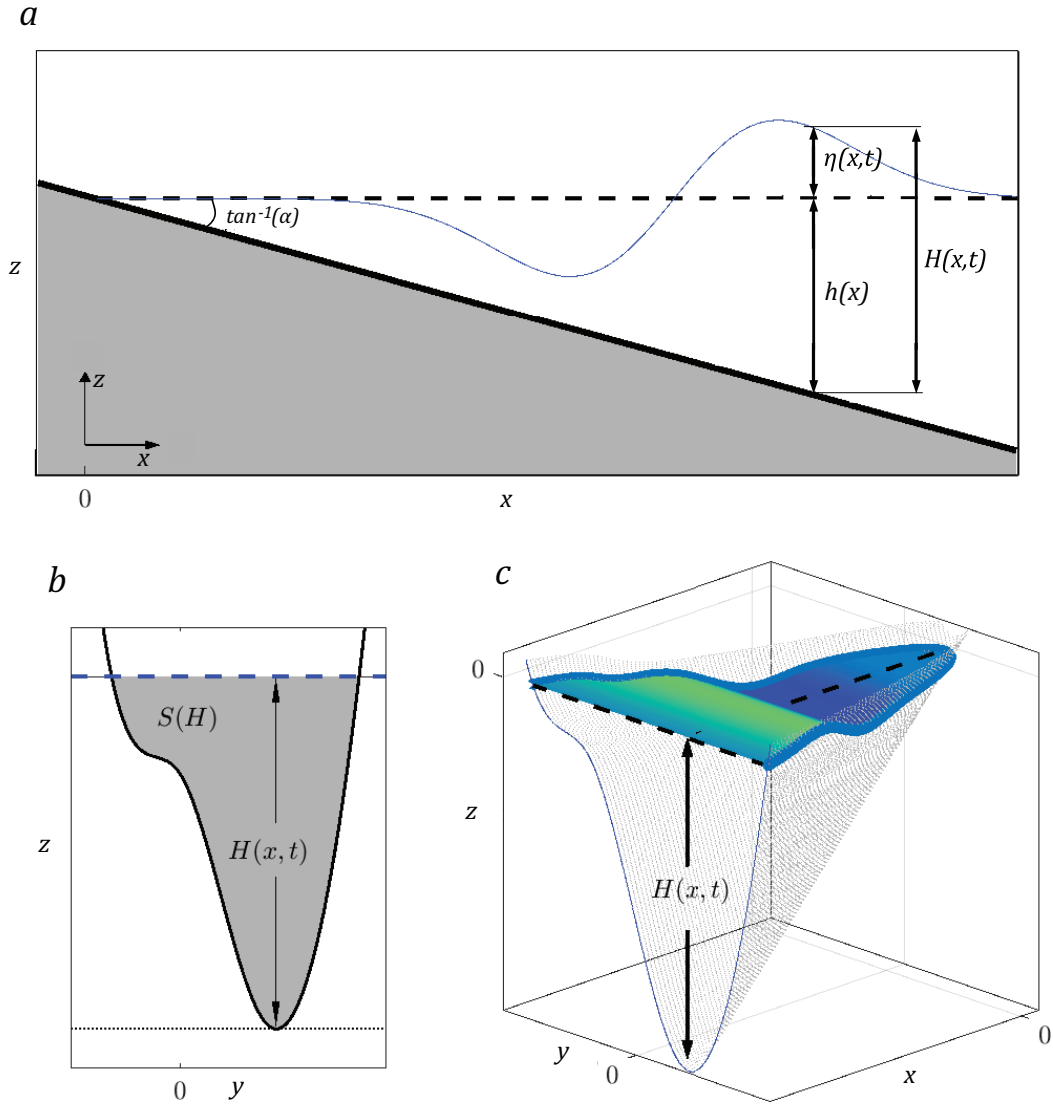}\caption{A: An $xz$ cross-section along the main axis of the bay. Both the unperturbed $h(x)$ (dashed black) and perturbed $H(x,t)$ (solid blue) water levels are displayed. B: $yz$ cross-section of a generic bay. Both $H(x,t)$ (dashed blue) and the cross-sectional area $S(H)$ (shaded area) are displayed. C: a 3-D view of the uniformly sloping bay, which cross-section is displayed in plot B.}\label{fig:mainbay}
\end{figure}

We make five additional assumptions: the wave is long (i.e. the height/length
ratio is small), friction and dispersion are both negligible, the bathymetry
(see Figure \ref{fig:mainbay}a) has the main axis located along $x$ and is
uniformly inclined. The SWE then reduce further to the 1+1 system (also called
shallow water)\citep{Stoker57,Lannes13}, which in dimension units and standard
notation reads
\begin{equation}
\left\{
\begin{array}
[c]{cccc}
\partial_{t}S+\partial_{x}(Su) & = & 0 & \text{(continuity equation)}\\
\partial_{t}u+u\partial_{x}u+g\partial_{x}\eta & = & 0 & \text{(momentum
equation)}
\end{array}
\right., \label{SWE}
\end{equation}
where:

\begin{itemize}
\item $\eta\left(  x,t\right)  $ is the water elevation over unperturbed water
level $z=0$. It need not be sign definite (can be positive or negative).

\item $u\left(  x,t\right)  $ is the flow velocity averaged over the
cross-section. Since the positive $x$-axis is directed off-shore, $u<0$ and
$u>0$ corresponds respectively to an in-coming wave (i.e. moving towards the
shore) and out-going wave (i.e. moving from the shore).

\item $S(x,t)$ is the cross-section area corresponding to the total water depth $H(x,t)=h\left(  x\right)  +\eta\left(  x,t\right)  $
along the main axis $x$ (see Figure \ref{fig:mainbay}b). Here $h(x)$ is the
distance between the hard bottom given by $z=-h\left(  x\right)  $ (along the
main axis $x$) and the unperturbed water level $z=0$. Note that $h\left(
x\right)  <0$ if $x<0$, and the shoreline location $x_s$ at the head of the bay is defined by $H(x_s,t)=h(x_s)+\eta(x_s,t)=0$.

We have assumed that our bathymetry (bay for short) is inclined. We agree to call a bay \emph{inclined} if
\begin{equation}
h\left(  x\right)  =\alpha x,\ \alpha>0,\ S(x,t)=S\left(  H(x,t)\right)
,\ \ dS/dH>0. \label{S}
\end{equation}
In other words, it has a constant slope and $S(x,t)$ depends on $(x,t)$ only
via $H(x,t)$. This is the case when the equation for the bottom is given by
\begin{equation}
z\left(  x,y\right)  =-\alpha x+f\left(  y\right)  \label{shoar line}
\end{equation}
with some $f\geq0$. Clearly, for a bay with one main axis $f\left(  y\right)
=0$ if and only if $y=0$. The unperturbed shoreline along the bay is given by $f\left(
y\right)  -\alpha x=0$.

\item $g$ is the acceleration due to gravity.
\end{itemize}

Note that while our bay is seemingly 3-D, its geometry is described by only
one single variable function $S\left(  H\right)  $ (or $f\left(  y\right)  $).
That is why the system (\ref{SWE}) for two unknown functions $\eta$ and $u$ is
essentially 1+1.

In this paper we are concerned with the initial boundary value problem (IBVP)
for (\ref{SWE}). That is, both $\eta$ and $u$ are specified at the initial
instant of time:
\begin{equation}
\eta(x,0)=\eta_{0}(x),~~~~~~u(x,0)=u_{0}(x).\label{IC}
\end{equation}
We will refer to such initial conditions (IC) as standard IC. Such
IC\ naturally occur, among others, in the study of tsunami waves generated by
landslides and near shore earthquakes. As a boundary condition (BC) we take
\begin{equation}
\eta(l,t)=\eta_b(t),~~~~~~u(l,t)=u_b(t), \label{BC}
\end{equation}
where $l$ is a fixed point $l>0$. Such BC appear e.g. in the study of finite
bathymetries and piece-wise inclined bays (see below). Typically,  $\eta_b(t)$ and $u_b(t)$ cannot be set up independently. As a matter of fact, it is already the case for subcritical flows ($u^2<gH$). We refer the interested reader to \citet{Antuono07,Antuono10} for extensive discussions on how to set up ``correct'' BC in this case and algorithms for solving IBVP for (\ref{SWE}) based on perturbation technics. Note that the assumption that the flow is subcritical is hard to enforce at the dry/wet boundary. Our considerations, on the other hand, do not need this assumption. The price to pay is that we need both $\eta_b(t)$ and $u_b(t)$. However, this set-up is not unrealistic as it would correspond, e.g., to the problem of computing the wave in a bay by measuring (in real time) the water displacement and velocity flow at a fixed point $x=b$.

It is convenient to go in (\ref{SWE}) over to dimensionless units
$\widetilde{x},\widetilde{t},\widetilde{\eta},\widetilde{u}$ defined from
\begin{equation}
x=\left(  H_{0}/\alpha\right)  \ \widetilde{x},\ t=\sqrt{H_{0}/g}
\ \widetilde{t}/\alpha,\ \eta=H_{0}\widetilde{\eta},\ \ u=\sqrt{H_{0}
g}\ \widetilde{u},\label{units}
\end{equation}
where $H_{0}$ is a typical (characteristic) height. Substituting (\ref{units})
into (\ref{SWE})-(\ref{IC}) and rewriting the first equation in (\ref{SWE})
using (\ref{S}), we have (omitting the tilde)
\begin{equation}
\left\{
\begin{array}
[c]{ccc}
\partial_{t}\eta+\left(  1+\partial_{x}\eta\right)  u+c^{2}\left(
x+\eta\right)  \partial_{x}u & = & 0\\
\partial_{t}u+u\partial_{x}u+\partial_{x}\eta & = & 0\\
\eta\left(  x,0\right)   & = & \eta_{0}\left(  x\right)  \\
u\left(  x,0\right)   & = & u_{0}\left(  x\right)  \\
\eta(l,t) & = & \eta_b(t)\\
u(l,t) & = & u_b(t)
\end{array}
\right.  ,\label{SWE1}
\end{equation}
where
\begin{equation}
c^{2}\left(  x\right)  :=S\left(  H_{0}x\right)  /H_{0}S^{\prime}\left(
H_{0}x\right)  \geq0.\label{c}
\end{equation}

The classical example of such idealized bathymetry is the plane infinite beach
(i.e. extending along the $y$ axis infinitely far). In this case, as one can
easily see, (\ref{SWE1}) takes the specifically simple form
\begin{equation}
\left\{
\begin{array}
[c]{ccc}
\partial_{t}\eta+\partial_{x}\left[  \left(  x+\eta\right)  u\right]  & = &
0\\
\partial_{t}u+u\partial_{x}u+\partial_{x}\eta & = & 0
\end{array}
\right.  . \label{SWE classical}
\end{equation}
The system (\ref{SWE classical}) has a quadratic nonlinearity. What is
remarkable about it is that the substitution
\begin{subequations}
\begin{align}
&
\begin{array}
[c]{ccc}
\varphi\left(  \sigma,\tau\right)  =u\left(  x,t\right)  , & \psi\left(
\sigma,\tau\right)  =\eta\left(  x,t\right)  +u^{2}\left(  x,t\right)  /2, &
\text{new unknowns},
\end{array}
\ \ \ \ \ \ \ \ \ \ \ \label{CG1}\\
&
\begin{array}
[c]{ccc}
\sigma=x+\eta\left(  x,t\right)  , & \ \tau=t-u\left(  x,t\right)  , &
\text{new variables,}
\end{array}\label{CG2}
\end{align}
turns it into
\end{subequations}
\begin{equation}
\left\{
\begin{array}
[c]{ccc}
\partial_{\tau}\psi+\sigma\partial_{\sigma}\varphi+\varphi & = & 0\\
\partial_{\tau}\varphi+\partial_{\sigma}\psi & = & 0
\end{array}
\right.  , \label{SWE linearized}
\end{equation}
which is a linear hyperbolic system! This substitution (in a slightly different form) was
introduced in the seminal paper by \cite{Carrier1} and is now referred to as
the Carrier-Greenspan (CG) transform. The form (\ref{CG1})-(\ref{CG2}) is
taken from \citet{Tuck72}. The system is typically written as one equation
\begin{equation}
\partial_{\tau}^{2}\psi=\ \sigma\partial_{\sigma}^{2}\psi+\ \partial_{\sigma
}\psi\text{ (or }\partial_{\tau}^{2}\varphi=\sigma\partial_{\sigma}^{2}
\varphi+2\partial_{\sigma}\varphi\text{),} \label{wave eqs}
\end{equation}
which is the wave equation with variable coefficients (also know in
mathematical physics as Klein-Gordon equation). Observe that $\left(
\sigma,\tau\right)  $ can be viewed as a hodograph plane and thus,
conceptually, the CG transform is a hodograph type transform (also called the
Carrier-Greenspan hodograph) that turns the nonlinear SWE (\ref{SWE1}) into
the linear wave equation (\ref{wave eqs}), which can, in turn, be explicitly
solved by the Hankel transform techniques for a variety of waveforms. This way
both boundary value \citep[e.g.][]{Synolakis87,Antuono07,Antuono10} and
initial value problems (IVP) \citep[e.g.][]{Yeh03,KANOGLU04,Kanoglu06} have
been extensively analyzed. A tremendous amount of information about the SWE
was learned this way (in particular, the important nonlinear process of the
run-up and run-down of long waves on the coast).

Same approach can be applied to more complicated inclined bays $S\left(
H\right)  $ \citep{Zahibo05}. For $f\left(  y\right)  \sim y^{2}$ (parabolic
bays) equation (\ref{wave eqs}) is the standard (constant coefficient) 1+1
wave equation and hence can be solved by the d'Alembert formula \citep{Didem}.
For an arbitrary power bay ($f\left(  y\right)  \sim\left\vert y\right\vert
^{m}$, $m>0$, which corresponds to $S(H)\sim H^{\left(  m+1\right)  /m}$ )
there is no d'Alembert solution but a similar to (\ref{wave eqs}) equation
takes place, which can be solved by the very same techniques \citep{MHarris16}
as for the plane beach. Note that the plane beach corresponds to $m=\infty$.
The case $m<1$ exhibits a new striking phenomenon: there may be more than one run-up/run-down.

More recently, the CG transform was generalized to inclined bathymetries of
arbitrary cross-section \citep{Rybkin}. In \citet{Raz2017} we finally show
that the very same substitution (\ref{CG1})-(\ref{CG2}) brings (\ref{SWE}) to
the linear system
\begin{equation}
\left\{
\begin{array}
[c]{ccc}
\partial_{\tau}\psi+c^{2}\left(  \sigma\right)  \partial_{\sigma}
\varphi+\varphi & = & 0\\
\partial_{\tau}\varphi+\partial_{\sigma}\psi & = & 0
\end{array}
\right.  , \label{SWE linearized S}
\end{equation}
where $c(\sigma)$ solely encodes the information about the shape of our bay.
The system (\ref{SWE linearized S}) easily implies
\begin{equation}
\partial_{\tau}^{2}\psi=c^{2}\left(  \sigma\right)  \partial_{\sigma}^{2}
\psi\ +\partial_{\sigma}\psi. \label{KG}
\end{equation}
Thus, surprisingly enough, the transformation (\ref{CG1})-(\ref{CG2}) is
universal for all inclined bathymetries and on the hodograph plane
(\ref{SWE linearized}) and (\ref{SWE linearized S}) (or (\ref{wave eqs}) and
(\ref{KG})) differ by the speed of propagation $c\left(  \sigma\right)  $
only. For power-shaped bays $S(H)\sim H^{\left(  m+1\right)  /m}$ we
immediately have $c^{2}\left(  \sigma\right)  =\dfrac{m}{m+1}\sigma$. In
particular, if $m=\infty$ (plane beach) then $c^{2}\left(  \sigma\right)
=\sigma$ and (\ref{KG}) turns into (\ref{wave eqs}) as expected. Only for
power bays can (\ref{KG}) be solved in terms of special functions. For all
other shapes (\ref{KG}) can effectively be solved and analyzed numerically.
See \citet{Harris15,Harris15a} and \citet{Raz2017} where detailed analysis is
done for trapezoidal, L, W, and other shapes. We also refer to
\citet{Anderson} for some extensions to piece-wise inclined power bays.

We emphasize that in the original SWE (\ref{SWE1}) the shoreline $x_s$ is moving
and it is the main problem with its analysis. On the hodograph plane that
point corresponds to the fixed point $\sigma=0$. Note that $c^{2}\left(
0\right)  =0$ and hence the differential operation on the right hand side of
(\ref{KG}) is singular. It is not a real issue from the mathematical point of
view but rather a strong manifestation of nonlinear effects of
run-up/run-down. This is one of the main (if not the main) advantages of the
CG transform.

The CG transform however has some serious drawbacks. For the reader's
convenience we explain in some detail what the issue is. Note first, that the
independent variable $\left(  \sigma,\tau\right)  $ defined by (\ref{CG2})
depend on the independent variable $\left(  \varphi,\psi\right)  $ defined by
(\ref{CG1}). This circumstance would not be an issue if the system (\ref{CG1})
was linear. But the second equation in (\ref{CG1}) has a quadratic
nonlinearity and this is the real problem (which, on the bright side, gives
fodder for extensive research). The reason is that the IC (\ref{IC}) on the
hodograph plane is no longer standard. As \citet{Johnson} simply puts it,
\textquotedblleft interchanging the dependent and independent variables
simplifies the governing equations, but complicates the boundary/initial
conditions." Indeed, under the transformation (\ref{CG2}), the (horizontal)
line $t=0$ in the physical plane $\left(  x,t\right)  $ becomes the parametric
curve $\Gamma=\left(  x+\eta_{0}(x),-u_{0}(x)\right)  $ in the $(\sigma,\tau)$
plane. From the first equation in (\ref{CG2}) one has $x=\gamma\left(
\sigma\right)  $, where $\gamma$ is the inverse function of $x+\eta_{0}(x)$
(i.e. solves the equation $x+\eta_{0}(x)=\sigma$). Thus
\begin{equation}
\Gamma=\left(
\begin{array}
[c]{c}
\sigma\\
-u_{0}|_{\gamma\left(  \sigma\right)  }
\end{array}
\right)  ,\ \sigma\geq0 \label{Gamma}
\end{equation}
is the curve in the hodograph plane where the IC\ are specified. From
(\ref{CG1}) we immediately have the transformed IC
\begin{equation}
\left.  \left(
\begin{array}
[c]{c}
\varphi\\
\psi
\end{array}
\right)  \right\vert _{\Gamma}=\left.  \left(
\begin{array}
[c]{c}
u_{0}\\
\eta_{0}+u_{0}^{2}/2
\end{array}
\right)  \right\vert _{\gamma\left(  \sigma\right)  },\ \sigma\geq0.
\label{IC transformed}
\end{equation}

\bigskip Similarly, the BC (\ref{BC}) transform as follows. Let $\gamma_b\left(
\tau\right)  $ be the inverse function of $\tau\left(  t\right)
=t-u_b\left(  t\right)  $ and
\begin{equation}
\Gamma_b=\left(
\begin{array}
[c]{c}
l+\eta_b|_{\gamma_b\left(  \tau\right)  }\\
\tau
\end{array}
\right)  ,\ \gamma_b\left(  \tau\right)  \geq0.\label{gamma 0}
\end{equation}
Then the BC in the hodograph plane are
\begin{equation}
\left.  \left(
\begin{array}
[c]{c}
\varphi\\
\psi
\end{array}
\right)  \right\vert _{\Gamma_b}=\left.  \left(
\begin{array}
[c]{c}
u_b\\
\eta_b+u_b^{2}/2
\end{array}
\right)  \right\vert _{\gamma_b\left(  \tau\right)  },\ \gamma_b\left(
\tau\right)  \geq0.\label{BC transformed}
\end{equation}
We can now see from (\ref{Gamma}) that $\Gamma$ is a horizontal line $\left(
\sigma,0\right)  $ if and only if the initial velocity $u_{0}=0.$ While the
latter is an important case, it is also quite restrictive, as we cannot always
assume that a tsunami wave is standing sill at the initial instant of time! If
$u_{0}\not =0$ then the IC (\ref{IC transformed}) is no longer standard and
the system (\ref{SWE linearized S}) (and hence the original SWE (\ref{SWE1}))
cannot be solved in closed form. Same issue of course takes place with
(\ref{gamma 0}): $\Gamma_b$ is a vertical line if and only if $\eta
_b\left(  \gamma_b\left(  \tau\right)  \right)  =const.$ These issues were
already noticed in \citep{Carrier1} and since then it has been a good open
problem how to make the CG transform run for general IVP. The main problem is
that the curves $\Gamma$ and $\Gamma_b$ also depend on IC $\left(  \eta
_{0},u_{0}\right)  $ and BC $\left(  \eta_b,u_b\right)  $. This problem
has drawn much attention but still only partial answers under various assumptions of the relative smallness of the IC
\citep[e.g.][]{Yeh03,Kanoglu06,Antuono07,Antuono10} are available and only in
the context of the plane beach. We will discuss these papers in some detail in
the main body of the text.

In the current paper, we put forward a complete solution to this problem for
arbitrary inclined bays. Our approach goes as follows\footnote{We outline the
main results here in Introduction. The derivations are given in the main
text.}. By the recipes discussed above, reduce the (nonlinear) SWE problem
(\ref{SWE1}) to the linear system (\ref{SWE linearized S}) with IC
(\ref{IC transformed}) and BC (\ref{BC transformed}) and write it in matrix
form
\begin{equation}
\left\{
\begin{array}
[c]{c}
\partial_{\tau}\Phi+A(\sigma)\partial_{\sigma}\Phi+B\Phi=0\\
\left.  \Phi\right\vert _{\Gamma}=\Phi_{0}\left(  \sigma\right)  \\
\left.  \Phi\right\vert _{\Gamma_b}=\Psi_{0}\left(  \tau\right)
\end{array}
\right.  ,\label{SWE in matrix form}
\end{equation}
where
\begin{align}
\ \ A\left(  \sigma\right)   &  =\left(
\begin{array}
[c]{cc}
0 & 1\\
c^{2}\left(  \sigma\right)   & 0
\end{array}
\right)  ,\ \ B=\left(
\begin{array}
[c]{cc}
0 & 0\\
1 & 0
\end{array}
\right)  ,\ \label{A, B, etc.}\\
\Phi &  =\left(
\begin{array}
[c]{c}
\varphi\\
\psi
\end{array}
\right)  ,\ \Phi_{0}\left(  \sigma\right)  =\left(
\begin{array}
[c]{c}
\varphi_{0}\left(  \sigma\right)  \\
\psi_{0}\left(  \sigma\right)
\end{array}
\right)  ,\ \Psi_{0}\left(  \tau\right)  =\left(
\begin{array}
[c]{c}
\varphi_b\left(  \tau\right)  \\
\psi_b\left(  \tau\right)
\end{array}
\right)  \nonumber
\end{align}
and
\[
\begin{array}
[c]{cc}
\varphi_{0}\left(  \sigma\right)  :=u_{0}\left(  \gamma\left(  \sigma\right)
\right)  , & \psi_{0}\left(  \sigma\right)  \ :=\eta_{0}\left(  \gamma\left(
\sigma\right)  \right)  +\varphi_{0}^{2}\left(  \sigma\right)  /2\\
\varphi_b\left(  \tau\right)  :=u_b\left(  \gamma_b\left(  \tau\right)
\right)  , & \psi_b\left(  \tau\right)  \ :=\eta_b\left(  \gamma
_b\left(  \tau\right)  \right)  +\varphi_b^{2}\left(  \tau\right)  /2
\end{array}
.
\]
Given accuracy $\varepsilon$ (could be arbitrarily small), we apply our
\emph{method of data projection,} put forward first in our recent
\citet{Nicolsky18} to find new standard IC $\left.  \Phi\right\vert _{\tau
=0}=\Phi_{n}\left(  \sigma\right)  $ given explicitly by
\begin{equation}
\Phi_{n}=\Phi_{0}+\sum_{k=1}^{n}\frac{1}{k!}\varphi_{0}^{k}\left(
D^{-1}\Delta\right)  ^{k}\Phi_{0},\label{Phi sub n}
\end{equation}
where $n$ is chosen to satisfy the accuracy $\varepsilon,$ and ($I$ is a 2x2
unit matrix)
\begin{equation}
D\left(  \sigma\right)  =I+\varphi_{0}^{\prime}\left(  \sigma\right)  A\left(
\sigma\right)  ,\ \Delta=-A\left(  \sigma\right)  \frac{d}{d\sigma
}-B.\label{D and delta}
\end{equation}
We call $\Phi_{n}$ \emph{the }$n^{\text{th}}$\emph{ projection of IC defined
on a curve onto the real line. }In a similar fashion we find $\Psi_{n}$,
projections of BC on some vertical line $\left(  \sigma_{0},\tau\right)$, e.g. $\sigma_0=l+\eta_0(l,0)$ to be compatible with the IC. Note that $n$ in the projections of IC and BC need not be the same. One can
now find the solution $\widetilde{\Phi}$ of the standard IVP
\[
\left\{
\begin{array}
[c]{c}
\partial_{\tau}\Phi+A(\sigma)\partial_{\sigma}\Phi+B\Phi=0\\
\left.  \Phi\right\vert _{\tau=0}=\Phi_{n}\left(  \sigma\right)  \\
\left.  \Psi\right\vert _{\sigma=\sigma_{0}}=\Psi_{n}\left(  \tau\right)
\end{array}
\right.  ,
\]
by any suitable method. Performing the inverse CG transform solves the
original problem (\ref{SWE1}) in the physical space. The latter is, in
general, not explicit but can easily be done numerically without affecting the
total accuracy, which remains $O\left(  \varepsilon\right)  .$ In fact, we can
call our method exact as the error it introduces can be made negligible
comparing with the one inherited by the shallow water approximation leading to
the very SWE (\ref{SWE1}).

Loosely speaking, the idea behind our method is to replace the IVP we cannot
solve with an equivalent one we can. We however emphasize that our equivalent
IC/BC would be very hard to guess. The reader is invited to amuse him/herself
with trying to unzip (\ref{Phi sub n}) even for $n=1$. It was the matrix form
(\ref{SWE in matrix form}) that made our derivation quite transparent.

Extensive numerical verification and simulations in Section \ref{Numerics}
show that our method is very robust and can be effectively used for rapid
forecasting of characteristics of the inundation zone. We will try to make our paper as self-contained as possible.

\section{\label{Data projection}\bigskip The method of data projection}

\bigskip In this section we introduce our\emph{ method of data projection} in
independent terms and most general situation (e.g. $(x,t)$ are not as in SWE
(\ref{SWE}) but rather $\left(  \sigma,\tau\right)  $, etc.). \ We consider
projections for IC and BC separately.

\subsection{Initial value problem}

Consider the hyperbolic system
\begin{equation}
\partial_{t}U=A\left(x\right)\partial_{x}U+B\left(x\right)U,\label{WE}
\end{equation}
where $U\left(  x,t\right)  $ is an $m$ column of dependent variables and
$A\left(  x\right)  $ and $B\left(  x\right)  $ are $m\times m$ matrices
independent of $t$. The domain for $\left(  x,t\right)  $ is inessential for
our consideration. Let $U$ be specified on some curve
\begin{equation}
\Gamma=\{(x,\tau(x))\}\label{curve}
\end{equation}
in the $\left(  x,t\right)  $ domain. Set up the following IVP
\begin{equation}
\left\{
\begin{array}
[c]{c}
\partial_{t}U=A\left(  x\right)  \partial_{x}U+B\left(  x\right)  U\\
U|_{\Gamma}=U_{0}\left(  x\right)
\end{array}
\right.  ,\label{IVP}
\end{equation}
where $U_{0}(x)$ is a \thinspace known function.

Note that we have not imposed any boundary conditions (BC) as we do not
actually intend to solve (\ref{IVP}) in this section. Thus we assume that
(\ref{IVP}) is supplemented by suitable BC. Conditions on $\tau(x)$ and
$U_{0}(x)$ will be given later.

If $\tau(x)=0$ then (\ref{IVP}) becomes the standard IVP solvable by a variety of classical techniques which
all break down if $\tau(x)\neq0$. Our idea is, given accuracy $\varepsilon$,
find standard IC $U|_{t=0}=\widetilde{U}_{0}$ such that the solution
$\widetilde{U}$ to
\begin{equation}
\left\{
\begin{array}
[c]{c}
\partial_{t}U=A\partial_{x}U+BU\\
U|_{t=0}=\widetilde{U}_{0}\left(  x\right)
\end{array}
\right.  \label{newIVP}
\end{equation}
would be within $O\left(  \varepsilon\right)  $ from the actual solution to
(\ref{IVP}) for all $\left(  x,t\right)  $ in the domain of interest. I.e. the
IVP (\ref{IVP}) and (\ref{newIVP}) are equivalent up to $O\left(
\varepsilon\right)  $.

We call the map $U_{0}\longrightarrow$ $\widetilde{U}_{0}$ the\emph{
projection of the data }$U_{0}=U|_{\Gamma}$\emph{ onto the real line}. The
reason why we can call it projection will be clear below from figure
\ref{fig:gammaproject}.

\begin{remark}
It is important feature of our method of data projection, that by the very construction both $U$ and $\widetilde{U}$ solve (exactly) the same equation (\ref{WE}) but satisfy different (equivalent) IC conditions. Of course $U$ and $\widetilde{U}$ can be made as close as one wishes (while $U_{0}$ and $\widetilde{U}_{0}$ need not be close at all). \
\end{remark}

To construct $\widetilde{U}_{0}$ we start out with applying the Taylor formula in one variable $t$ to the
solution (still unknown) $U\left(  x,t\right)  $ of (\ref{WE}). For each fixed
point $(x,t)$ we then have
\begin{equation}
U(x,0)=\sum_{k=0}^{n}\frac{1}{k!}\partial_{t}^{k}U(x,t)(-t)^{k}+E_{n}\left(
x,t\right)  \label{taylor}
\end{equation}
with some error $E_{n}$. I.e., we fix $\left(  x,t\right)  $ and apply the
Taylor formula to the point $\left(  x,0\right)  $ and not the other way
around. Note that the right hand side of (\ref{taylor}) is independent of $t$
(because $t=0$ in left hand side). Taking in (\ref{taylor}) $t=\tau\left(
x\right)  $ yields
\[
U|_{t=0}\left(  x\right)  =\sum_{k=0}^{n}\frac{(-\tau(x))^{k}}{k!}\left.
\left[  \partial_{t}^{k}U(x,t)\right]  \right\vert _{\Gamma}+\left.
E_{n}\right\vert _{\Gamma}.
\]
Introduce

\begin{equation}
U_{n}(x):=\sum_{k=0}^{n}\frac{(-\tau(x))^{k}}{k!}\left.  \left[  \partial
_{t}^{k}U(x,t)\right]  \right\vert _{\Gamma}, \label{nth order projection}
\end{equation}
which we call \emph{the }$n^{\text{th}}$\emph{ order projection} of initial
data $\left.  U\right\vert _{\Gamma}$ onto the real line. We can now claim
that if we are able to compute all $\left.  \left[  \partial_{t}
^{k}U(x,t)\right]  \right\vert _{\Gamma}$ via $U_{0}=U|_{\Gamma}$ and $A$, $B$
then $U_{n}$ produces a desirable standard IC $\widetilde{U}_{0}$ from
(\ref{newIVP}). Indeed, given error $\varepsilon$ (no matter how small), we
take $n$ so large as $\left\vert E_{n}\right\vert <\varepsilon$ everywhere in
the domain of interest for $\left(  x,t\right)  $ and hence
\[
U|_{t=0}=U_{n}+O\left(  \varepsilon\right)  .
\]
Thus the solution $\widetilde{U}$ to (\ref{newIVP}) with $\widetilde{U}
_{0}=U_{n}$ will coincide with the solution $U$ of (\ref{IVP}) up to $O\left(
\varepsilon\right)  $.

So, it remains to compute the Taylor coefficients in
(\ref{nth order projection}). The zeroth one is obvious
\[
U_{0}(x)=U(x,t)|_{\Gamma}
\]
and it is the data in (\ref{IVP}). We call it the $0^{\text{th}}$ order
projection of the data $U_{0}=U|_{\Gamma}$ onto the real line. All other
Taylor coefficients in (\ref{nth order projection}) can also be explicitly
computed. Start with the first one. Restricting (\ref{WE}) to $\Gamma$,
suppressing the variable, and introducing the convenient short-hand notation
\[
\Delta U:=\left(  A\partial_{x}+B\right)  U,
\]
we have
\begin{equation}
\left.  (\partial_{t}U)\right\vert _{\Gamma}=\left.  (\Delta U)\right\vert
_{\Gamma}=A\left.  (\partial_{x}U)\right\vert _{\Gamma}+\left.  BU\right\vert
_{\Gamma}.\label{1st coeff}
\end{equation}
Compute now $\left.  (\partial_{x}U)\right\vert _{\Gamma}$. To avoid possible confusion, note that $(\partial_{x}U)|_{\Gamma}\not =
\frac{d}{dx}\left(  U|_{\Gamma}\right)  $ (indeed, $\left(  \partial
_{x}U\right)  |_{\Gamma}=\left(  \partial_{x}U\right)  |_{t=\tau\left(
x\right)  }$ whereas $\frac{d}{dx}\left(  U|_{\Gamma}\right)  =\frac{d}
{dx}U\left(  x,\tau\left(  x\right)  \right)  $). By the chain rule
(prime denotes $d/dx$)
\begin{align*}
\frac{d}{dx}(U|_{\Gamma}) &  =\frac{d}{dx}U(x,\tau(x))\\
&  =(\partial_{x}U)|_{\Gamma}+(\partial_{t}U)\ \tau^{\prime}\qquad\text{(by
the chain rule)}\\
&  =(\partial_{x}U)|_{\Gamma}+\left\{  A(\partial_{x}U)|_{\Gamma}+BU|_{\Gamma
}\right\}  \ \tau^{\prime}\text{ \ (by (\ref{1st coeff}))}\\
&  =\left(  I+\tau^{\prime}A\right)  \left(  \partial_{x}U\right)  |_{\Gamma
}+\tau^{\prime}BU|_{\Gamma}\\
&  =D\left(  \partial_{x}U\right)  |_{\Gamma}+\tau^{\prime}BU|_{\Gamma},
\end{align*}
where $I$ is the unit matrix and
\[
D:=I+\tau^{\prime}A.
\]
Thus,
\[
\frac{d}{dx}(U|_{\Gamma})=D\left(  \partial_{x}U\right)  |_{\Gamma}+\tau
^{\prime}BU|_{\Gamma}
\]
and hence, solving this equation for $\left(  \partial_{x}U\right)  |_{\Gamma
}$, we have
\begin{align*}
\left(  \partial_{x}U\right)  |_{\Gamma} &  =D^{-1}\left(  (U|_{\Gamma
})^{\prime}-\tau^{\prime}BU|_{\Gamma}\right)  \\
&  =D^{-1}\left(  I\ d/dx-\tau^{\prime}B\right)  U|_{\Gamma}.
\end{align*}
Substituting this equation into (\ref{1st coeff}) yields
\begin{align*}
\left(  \partial_{t}U\right)  |_{\Gamma} &  =AD^{-1}\left(  I\ d/dx-\tau
^{\prime}B\right)  U|_{\Gamma}+BU|_{\Gamma}\\
&  =\left\{  AD^{-1}d/dx-\tau^{\prime}AD^{-1}B+B\right\}  U|_{\Gamma}\\
&  =\left\{  AD^{-1}d/dx+\left(  I-\tau^{\prime}A(I+\tau^{\prime}
A)^{-1}\right)  B\right\}  U|_{\Gamma}
\end{align*}
Here we have noticed that the matrices $\tau^{\prime}A$ and $(I+\tau^{\prime
}A)^{-1}$ commute and hence
\[
I-\tau^{\prime}A(I+\tau^{\prime}A)^{-1}=(I+\tau^{\prime}A)^{-1}=D^{-1}.
\]
Since $A$ and $D$ also commute,
\[
\left(  \partial_{t}U\right)  |_{\Gamma}=D^{-1}\left(  A\ d/dx+B\right)
U|_{\Gamma}=D^{-1}\Delta U_{0}
\]
and thus for the first order Taylor coefficient we finally have
\begin{equation}
\left(  \partial_{t}U\right)  |_{\Gamma}=D^{-1}\Delta\left(  U|_{\Gamma
}\right)  =D^{-1}\Delta U_{0}.\label{partial of t}
\end{equation}

Our computation of higher order Taylor coefficients will be based on the
following observation. Since $\partial_{t}$ and $A\left(  x\right)  $ commute,
$\partial_{t}U$ is also a solution to $\partial_{t}U=\Delta U$, i.e.
\[
\partial_{t}\left(  \partial_{t}U\right)  =\Delta\left(  \partial_{t}U\right)
,
\]
and on the curve
\[
\left(  \partial_{t}U\right)  |_{\Gamma}=D^{-1}\Delta U_{0}\equiv U_{1}
\qquad\text{(by (\ref{partial of t})).}
\]
Thus, if $U$ is the solution originated from $U_{0}$ then $\partial_{t}U$ is
the solution originated from the IC $U_{1}=D^{-1}\Delta U_{0}$. By induction
one concludes that $\partial_{t}^{k}U$ is the solution originated from
$U_{k}=D^{-1}\Delta U_{k-1}$, $k=2,3,4,...$\newline Therefore, we get the
following nice formula
\[
\left(  \partial_{t}^{k}U\right)  |_{\Gamma}=\left(  D^{-1}\Delta\right)
^{k}U_{0},\ \ k=0,1,2,...
\]

Substituting this into (\ref{nth order projection}) we finally arrive at

\begin{equation}
U_{n}(x)=\sum_{k=0}^{n}\frac{1}{k!}(-\tau(x))^{k}\left(  D^{-1}\Delta\right)
^{k}U_{0}(x), \label{final nth order projection}
\end{equation}
where, if we recall,
\begin{align}
&  D=I+\tau^{\prime}(x)A(x)\label{D}\\
&  \Delta=A(x)\ d/dx+B(x). \label{delta}
\end{align}
Note that in (\ref{delta}) we have the full derivative as in (\ref{final nth order projection}) we have only one variable. We indicate that $U_{0}\left(  x\right)  $, the original IC on the curve, is only the zero order approximation of our $U_{n}\left(  x\right)$, which suggests that $U_{n}\left(  x\right)  $ is rather a projection than an approximation.

Explicit expanding $\left(  D^{-1}\Delta\right)^{k}$ in (\ref{final nth order projection}) is extremely unwieldy but numerical implementation of (\ref{final nth order projection}) does not cause any problems. Similar to \citet{Nicolsky18}, where we have a bit more complicated formula for $U_{n}(x)$, the following recursion formula for (\ref{final nth order projection}) could be obtained:%
\begin{align*}
U_{n}  &  =U_{n-1}-\frac{1}{n}\tau D^{-1}A\left(  U_{n-1}^{\prime}-U_{n-2}^{\prime}\right) \\
&  +\frac{n-1}{n}\tau^{\prime}D^{-1}A\left(  U_{n-1}-U_{n-2}\right)  -\frac{1}{n}\tau D^{-1}B\left(U_{n-1}-U_{n-2}\right)  .
\end{align*}

It follows from (\ref{final nth order projection}) that the map $U_{0}\longrightarrow U_{n}$ is linear and well-defined as long as the matrix $D$ is non-singular, i.e.
\begin{equation}
\det\left(  I+\tau^{\prime}(x)A(x)\right)  \neq0,\label{det cond}
\end{equation}
and the entries of $A$, $B$, and $\tau^{\prime}$ are at least $n$ times continuously differentiable. We investigate these conditions for the power-shaped bays later in Subsection \ref{finite}.

\subsection{Boundary value problem\label{BVP general}}
The considerations of the previous subsection can be easily adjusted to the BVP. Let $U$ be a column of dependent variables specified on a curve $\Gamma=\{(f(t),t)|t\ge0\}$, i.e.
\begin{equation}
U|_{\Gamma}=U_{0}.\label{curve_bc}
\end{equation}
Condition (\ref{curve_bc}) can be viewed as a boundary condition at a variable
point $x_b=f(t)$. Such a situation occurs when we study the SWE on a finite interval $x_s \le x \le l=const$ and map it to the hodograph plane using the CG transform. While the shoreline $x_s$ becomes fixed in the hodograph plane, the other end becomes a floating point. As above we show that given accuracy $\varepsilon$, we can
find a standard boundary condition at some point $x_{0}$ such that the IBVP
problem,
\begin{equation}
\left\{
\begin{array}
[c]{c}
\partial_{t}U=A\partial_{x}U+BU\\
U|_{x=x_{0}}=\widetilde{U_{0}}(t)\\
\text{some IC}
\end{array}
\right.  \label{IBVP}
\end{equation}
has the solution $\widetilde{U_{0}}(x,t)$ different from the solution $U(x,t)$
to (\ref{WE})-(\ref{curve_bc}) by not more than $\varepsilon$.

Let $x_{0}$ be a fixed point, e.g. we can take $x_{0}=f(0)$. Then by Taylor's
formula we have
\begin{equation}
U(x_{0},t)=\sum_{k=0}^{n}\frac{1}{k!}\partial_{x}^{k}U(x,t)(x_{0}-x)^{k}+E_{n},\label{taylor form}
\end{equation}
where $x$ is taken so that $(x,t)\in\Gamma$ and $E_{n}$ is the error
term. Consequently, if we are able to find $\partial_{x}^{k}U(x,t)$ and demonstrate
that $|E_{n}|<\varepsilon$ then,
\[
\widetilde{U_{0}}(t)=\sum_{k=0}^{n}\frac{1}{k!}\partial_{x}^{k}U(x,t)(x_{0}
-x)^{k}
\]
will be the desired BC in (\ref{IBVP}). Thus the problem boils down again to
finding $\partial_{x}^{k}U|_{\Gamma}$ in terms of $U_{0}$ and $\Gamma$. Below
$U|_{\Gamma}=U(f(t),t)$ and of course $\frac{d}{dt}U|_{\Gamma}\neq
(\partial_{t}U)|_{\Gamma}$. Differentiating $U|_{\Gamma}=U_{0}(t)$ by the
chain rule

\[
\begin{split}
\frac{d}{dt}(U|_{\Gamma}) &  =\partial_{x}U|_{\Gamma}\cdot x^{\prime
}(t)+\partial_{t}U|_{\Gamma}\\
&  =\partial_{x}U|_{\Gamma}\cdot f^{\prime}+(A\partial_{x}U)|_{\Gamma
}+(BU)|_{\Gamma}\\
&  =(A|_{\Gamma}+f^{\prime}I)\partial_{x}U|_{\Gamma}+B|_{\Gamma}(U|_{\Gamma})
\end{split}
\]
we find
\[
\partial_{x}U|_{\Gamma}=\left(  A|_{\Gamma}+f^{\prime}I\right)  ^{-1}\left(
\frac{d}{dt}(U|_{\Gamma})-B|_{\Gamma}(U|_{\Gamma})\right)  .
\]
Thus, recalling that $U|_{\Gamma}=U_{0}$, we have
\begin{equation}
\partial_{x}U|_{\Gamma}=D^{-1}(U_{0}^{\prime}-B|_{\Gamma}U_{0}
)\label{partial x of U}
\end{equation}
where $D\equiv A|_{\Gamma}+f^{\prime}I$. At this point we make a simplifying
assumption pertinent to our specific equation below. Suppose that $A^{\prime
}=const$ and $B^{\prime}=0$. In such a case a nice formula can be derived.
Indeed, differentiating (\ref{WE}) with respect to $x$, we have
\[
\partial_{t}(\partial_{x}U)=A\partial_{x}(\partial_{x}U)+\left(
dA/dx+B\right)  \partial_{x}U.
\]
Thus, if $U$ solves (\ref{WE})-(\ref{curve_bc}) then $\partial_{x}U$ solves
\begin{equation}
\begin{array}
[c]{c}
\partial_{t}U=A\partial_{x}U+(B+A^{\prime})U\\
U|_{\Gamma}=U_{1}
\end{array}
,\label{partial x solves}
\end{equation}
where $U_{1}=D^{-1}\left(  U_{0}^{\prime}-BU_{0}\right)  $. One can now see
that the new problem (\ref{partial x solves}) is different from (\ref{WE}
)-(\ref{curve_bc}) by the substitutions
\begin{equation}
B\longrightarrow B_{1}=B+A^{\prime}, ~~~~~~~~~ U_{0}\longrightarrow U_{1}.
\end{equation}\label{sub}
This means that (\ref{partial x of U}) applies with updated data $B_{1}$ and
$U_{1}$:
\begin{align*}
\partial_{x}^{2}U|_{\Gamma} &  =D^{-1}(U_{1}^{\prime}-B_{1}U_{1})\\
&  =D^{-1}\left(  \frac{d}{dt}-B_{1}\right)  U_{1}\\
&  =D^{-1}\left(  \frac{d}{dt}-B_{1}\right)  D^{-1}\left(  \frac{d}
{dt}-B\right)  U_{0}
\end{align*}
and the following pattern emerges
\begin{equation}
\partial_{x}^{k}U|_{\Gamma}=D^{-1}\left(  \frac{d}{dt}-B_{k-1}\right)
D^{-1}\left(  \frac{d}{dt}-B_{k-2}\right)  ...D^{-1}\left(  \frac{d}{dt}
-B_{0}\right)  U_{0},\label{k partial x of U}
\end{equation}
where $B_{j}=B_{j-1}+A^{\prime},\,j=1,2,...,k$ with $B_{0}=B$.\newline Thus
problem (\ref{IBVP}) is completely solved. Indeed let $A$, $B$ be subject to
$A^{\prime}=const$, $B^{\prime}=0$. By taking $n$ large enough such that the
solution $U$ to
\[
\left\{
\begin{array}
[c]{c}
\partial_{t}U=A\partial_{x}U+BU\\
U|_{\Gamma}=U_{0}\\
\text{some}\,\text{\ IC}
\end{array}
\right.  ,\ \Gamma=\left\{  (f(t),t)\right\}
\]
differs by not more than $\varepsilon$ from $\widetilde{U}$ which solves the
standard IBVP
\[
\left\{
\begin{array}
[c]{c}
\partial_{t}U=A\partial_{x}U+BU\\
U|_{x=L}=\widetilde{U_{0}}(t)\\
\text{some IC}
\end{array}
\right.  ,
\]
where
\begin{equation}
\widetilde{U_{0}}(t)=\left.  \sum_{k=0}^{n}\frac{(x_{0}-f(t))^{k}}{k!}
\partial_{x}^{k}U(x,t)\right\vert _{\Gamma},\label{projection for BC}
\end{equation}
and $\partial_{x}^{k}U|_{\Gamma}$ is given by (\ref{k partial x of U}). The choice
of $x_{0}$ is at our disposal. It can be chosen to be compatible with the IC.

We conclude this section by noting that the Carrier-Greenspan transform reduces the SWE to a linear wave equation on a variable interval. Our method then makes this interval fixed and the transformed equation can then be effectively solved by any applicable method.

\section{Method of Data Projection for the SWE}
In this section we apply our method to the study of the SWE (\ref{SWE}) for inclined bays with arbitrary IC. The application of the data projection method for the arbitrary BC will be discussed in the following section in the context of power-shaped inclined bays, because of certain restrictions on the matrices. As we have seen, projections of IC and BC do not affect the equation itself and therefore we can do IC and BC separately, which is of course technically much easier. After that one can merely put them together by superposition. 

\subsection{\bigskip Conditions on the IC for the data projection method}

With all formulas prepared in the previous section we only need to show that
our machinery to solve (\ref{SWE1}) runs smoothly unless the gradient
catastrophe (wave breaking)\ occurs. The latter happens when invertibility of
the CG transform (\ref{CG2}) or its inverse fails \citep[see][]{Rybkin}, i.e.
when
\begin{equation}
\det\frac{\partial\left(  \sigma,\tau\right)  }{\partial\left(  x,t\right)
}=0\text{ or }\det\frac{\partial\left(  x,t\right)  }{\partial\left(
\sigma,\tau\right)  }=0. \label{detJ}
\end{equation}
The Jacobian in (\ref{detJ}) on the left has a nice formula
\begin{align*}
\det\frac{\partial\left(  \sigma,\tau\right)  }{\partial\left(  x,t\right)  }
&  =\det\left(
\begin{array}
[c]{cc}
\partial_{x}\sigma & \partial_{x}\tau\\
\partial_{t}\sigma & \partial_{t}\tau
\end{array}
\right)  =\partial_{x}\sigma\partial_{t}\tau-\partial_{t}\sigma\partial
_{x}\tau\\
&  =\det\left(
\begin{array}
[c]{cc}
1+\partial_{x}\eta & -\partial_{x}u\\
\partial_{t}\eta & 1-\partial_{t}u
\end{array}
\right)  \text{ \ \ (by (\ref{CG2})}\\
&  =(1+\partial_{x}\eta)^{2}-\left(  c(x+\eta)\partial_{x}u\right)  ^{2}\text{
(by (\ref{SWE1}))}
\end{align*}
Hence, at $t=0$
\begin{align*}
\left.  \det\frac{\partial\left(  \sigma,\tau\right)  }{\partial\left(
x,t\right)  }\right\vert _{t=0}  &  =\left(  1+\eta_{0}^{\prime}\right)
^{2}-\left(  c(x+\eta_{0})u_{0}^{\prime}\right)  ^{2}\text{ (in }x\text{
variable)}\\
&  =\left(  1+\eta_{0}^{\prime}|_{\gamma\left(  \sigma\right)  }\right)
^{2}-\left(  c(\sigma)u_{0}^{\prime}|_{\gamma\left(  \sigma\right)  }\right)
^{2}\text{ (in }\sigma\text{ variable).}
\end{align*}
Consequently, the condition for the CG transform invertibility reads (recall
$\gamma\left(  \sigma\right)  $ is the solution to $\sigma=x+\eta_{0}(x)$)
\begin{equation}
\left(  \left(  1+\partial_{x}\eta\right)  ^{2}-c^{2}(x+\eta)\left(
\partial_{x}u\right)  ^{2}\right)  ^{\pm1}\not =0\text{ (for all }\left(
x,t\right)  \text{),} \label{nonbreaking for all x,t}
\end{equation}
\begin{equation}
\left(  \left(  1+\eta_{0}^{\prime}\right)  ^{2}-\left(  c(x+\eta_{0}
)u_{0}^{\prime}\right)  ^{2}\right)  ^{\pm1}\not =0\text{ (for }t=0\text{),}
\label{nonbreaking cond}
\end{equation}
\begin{equation}
\left(  \left(  1+\eta_{0}^{\prime}|_{\gamma\left(  \sigma\right)  }\right)
^{2}-\left(  c(\sigma)u_{0}^{\prime}|_{\gamma\left(  \sigma\right)  }\right)
^{2}\right)  ^{\pm1}\not =0\text{ (for all }\sigma\geq0\text{).}
\label{nonbreaking cond1}
\end{equation}
Recall that (\ref{SWE1}) in matrix form in $\left(  \sigma,\tau\right)  $ is
\begin{equation}
\left\{
\begin{array}
[c]{c}
\partial_{\tau}\left(
\begin{array}
[c]{c}
\varphi\\
\psi
\end{array}
\right)  +\left(
\begin{array}
[c]{cc}
0 & 1\\
c^{2}\left(  \sigma\right)  & 0
\end{array}
\right)  \partial_{\sigma}\left(
\begin{array}
[c]{c}
\varphi\\
\psi
\end{array}
\right)  +\left(
\begin{array}
[c]{cc}
0 & 0\\
1 & 0
\end{array}
\right)  \left(
\begin{array}
[c]{c}
\varphi\\
\psi
\end{array}
\right)  =0\\
\left.  \left(
\begin{array}
[c]{c}
\varphi\\
\psi
\end{array}
\right)  \right\vert _{\Gamma}=\left.  \left(
\begin{array}
[c]{c}
u_{0}\\
\eta_{0}+u_{0}^{2}/2
\end{array}
\right)  \right\vert _{\gamma\left(  \sigma\right)  }
\end{array}
\right.  . \label{IVP for SWE in matrix form}
\end{equation}
It follows from the previous section that (\ref{IVP for SWE in matrix form})
can be solved by our method of data projection if the condition
(\ref{det cond}) holds. Rewriting (\ref{det cond}) for our specific
(\ref{IVP for SWE in matrix form}) yields
\begin{align}
\det D  &  =\det\left(
\begin{array}
[c]{cc}
1 & \left(  u_{0}|_{\gamma\left(  \sigma\right)  }\right)  ^{\prime}\\
c^{2}\left(  \sigma\right)  \left(  u_{0}|_{\gamma\left(  \sigma\right)
}\right)  ^{\prime} & 1
\end{array}
\right)  =1-\left[  c\left(  \sigma\right)  \left(  u_{0}|_{\gamma\left(
\sigma\right)  }\right)  ^{\prime}\right]  ^{2}\nonumber\\
&  =1-\left[  c\left(  \sigma\right)  \frac{u_{0}^{\prime}|_{\gamma\left(
\sigma\right)  }}{1+\eta_{0}^{\prime}|_{\gamma\left(  \sigma\right)  }
}\right]  ^{2}=\frac{\left(  1+\eta_{0}^{\prime}|_{\gamma\left(
\sigma\right)  }\right)  ^{2}-\left(  c\left(  \sigma\right)  u_{0}^{\prime
}|_{\gamma\left(  \sigma\right)  }\right)  ^{2}}{\left(  1+\eta_{0}^{\prime
}|_{\gamma\left(  \sigma\right)  }\right)  ^{2}}\nonumber\\
&  =\left(  1+\eta_{0}^{\prime}|_{\gamma\left(  \sigma\right)  }\right)
^{-2}\left.  \det\frac{\partial\left(  \sigma,\tau\right)  }{\partial\left(
x,t\right)  }\right\vert _{t=0}. \label{det com}
\end{align}
Here we have used
\[
\left(  u_{0}|_{\gamma\left(  \sigma\right)  }\right)  ^{\prime}=\frac
{u_{0}^{\prime}|_{\gamma\left(  \sigma\right)  }}{1+\eta_{0}^{\prime}
|_{\gamma\left(  \sigma\right)  }},
\]
which follows merely from the chain rule $\left(  u_{0}|_{\gamma\left(
\sigma\right)  }\right)  ^{\prime}=u_{0}^{\prime}|_{\gamma\left(
\sigma\right)  }\gamma^{\prime}\left(  \sigma\right)  $ and $\ \gamma^{\prime
}\left(  \sigma\right)  =\left(  1+\eta_{0}^{\prime}|_{\gamma\left(
\sigma\right)  }\right)  ^{-1}$.

It immediately follows from (\ref{det com}) that if the Jacobi matrix $\left.
\frac{\partial\left(  \sigma,\tau\right)  }{\partial\left(  x,t\right)
}\right\vert _{t=0}$ is nonsingular then so is $D$. Thus the condition
(\ref{nonbreaking cond1}) is sufficient for $D$ to be nonsingular.

\subsection{\label{algorithm}Algorithm of solving SWE with arbitrary IC}

\begin{figure}[ptb]
\centering
\includegraphics[width = 0.9\textwidth]{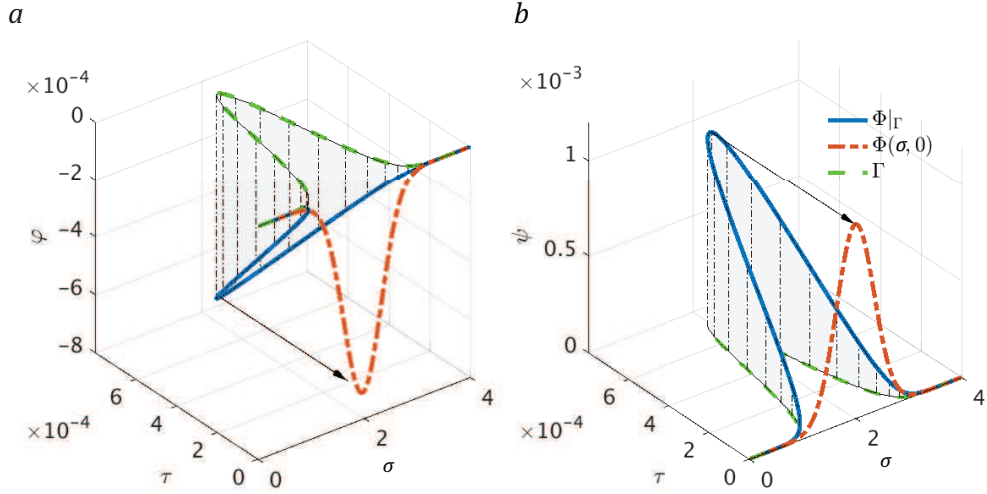}
\caption{Projection of components $\Phi|_{\Gamma}=(\varphi,\psi)$ onto the plane $\tau=0$ for the initial disturbance with the non-zero water velocity.}
\label{fig:gammaproject}
\end{figure}

For the reader's convenience we summarized here our main result putting
together all related formulas in one place.

Consider the IVP (\ref{SWE1}) (i.e. the BC are replaced with a natural
condition that $\eta$ and $u$ are both bounded) with non-breaking IC (i.e.
subject to (\ref{nonbreaking cond}). Perform the generalized CG transform
\[
\begin{array}
[c]{cc}
\varphi\left(  \sigma,\tau\right)  =u\left(  x,t\right)  , & \psi\left(
\sigma,\tau\right)  =\eta\left(  x,t\right)  +u^{2}\left(  x,t\right)  /2,\\
\sigma=x+\eta\left(  x,t\right)  , & \tau=t-u\left(  x,t\right)  ,
\end{array}
\]
which reduces (\ref{SWE1}) to the linear IVP (but with IC on a curve)
\begin{equation}
\left\{
\begin{array}
[c]{ccc}
\partial_{\tau}\psi+c^{2}\left(  \sigma\right)  \partial_{\sigma}
\varphi+\varphi & = & 0\\
\partial_{\tau}\varphi+\partial_{\sigma}\psi & = & 0\\
\varphi\left(  \sigma,-\varphi_{0}\left(  \sigma\right)  \right)   & = &
\varphi_{0}\left(  \sigma\right)  \\
\psi\left(  \sigma,-\varphi_{0}\left(  \sigma\right)  \right)   & = & \psi
_{0}\left(  \sigma\right)
\end{array}
\right.  ,\label{IVP on curve}
\end{equation}
where $\gamma\left(  \sigma\right)  $ is the inverse function of
$\sigma=x+\eta_{0}\left(  x\right)  $ and
\[
\varphi_{0}\left(  \sigma\right)  =u_{0}\left(  \gamma\left(  \sigma\right)
\right)  ,\ \ \psi_{0}\left(  \sigma\right)  \ =\eta_{0}\left(  \gamma\left(
\sigma\right)  \right)  +\varphi_{0}^{2}\left(  \sigma\right)  /2.
\]
Given accuracy $\varepsilon$ we replace (\ref{IVP on curve}) with the standard
IVP
\begin{equation}
\left\{
\begin{array}
[c]{ccc}
\partial_{\tau}\psi+c^{2}\left(  \sigma\right)  \partial_{\sigma}
\varphi+\varphi & = & 0\\
\partial_{\tau}\varphi+\partial_{\sigma}\psi & = & 0\\
\varphi\left(  \sigma,0\right)   & = & \varphi_{n}\left(  \sigma\right)  \\
\psi\left(  \sigma,0\right)   & = & \psi_{n}\left(  \sigma\right)
\end{array}
\right.  ,\label{standard IVP}
\end{equation}
where
\begin{equation}
\left(
\begin{array}
[c]{c}
\varphi_{n}\left(  \sigma\right)  \\
\psi_{n}\left(  \sigma\right)
\end{array}
\right)  =\left(
\begin{array}
[c]{c}
\varphi_{0}\left(  \sigma\right)  \\
\ \psi_{0}\left(  \sigma\right)
\end{array}
\right)  +\sum_{k=1}^{n}\frac{\varphi_{0}^{k}\left(  \sigma\right)  }
{k!}\left(  D^{-1}\Delta\right)  ^{k}\left(
\begin{array}
[c]{c}
\varphi_{0}\left(  \sigma\right)  \\
\ \psi_{0}\left(  \sigma\right)
\end{array}
\right)  ,\label{series}
\end{equation}
\[
D=\left(
\begin{array}
[c]{cc}
1 & \varphi_{0}^{\prime}\left(  \sigma\right)  \\
c^{2}\left(  \sigma\right)  \varphi_{0}^{\prime}\left(  \sigma\right)   & 1
\end{array}
\right)  ,\ \Delta=-\left(
\begin{array}
[c]{cc}
0 & 1\\
c^{2}\left(  \sigma\right)   & 0
\end{array}
\right)  \frac{d}{d\sigma}-\left(
\begin{array}
[c]{cc}
0 & 0\\
1 & 0
\end{array}
\right)  ,
\]
and $n$ is chosen so that\footnote{$\left\Vert \cdot\right\Vert $ stands for
the Eucledian norm.}
\[
\max_{\sigma\geq0}\left\Vert \frac{\varphi_{0}^{n+1}\left(  \sigma\right)
}{\left(  n+1\right)  !}\left(  D^{-1}\Delta\right)  ^{n+1}\left(
\begin{array}
[c]{c}
\varphi_{0}\left(  \sigma\right)  \\
\psi_{0}\left(  \sigma\right)
\end{array}
\right)  \right\Vert <\varepsilon.
\]
Solve (\ref{standard IVP}) analytically or numerically for $\left(
\varphi\left(  \sigma,\tau\right)  ,\psi\left(  \sigma,\tau\right)  \right)
$. This $\left(  \varphi,\psi\right)  $ also solves (\ref{IVP on curve}) up to
error $O\left(  \varepsilon\right)  $. Performing the inverse CG transform
\[
\begin{array}
[c]{cc}
u\left(  x,t\right)  =\varphi\left(  \sigma,\tau\right)  , & \eta\left(
x,t\right)  =\psi\left(  \sigma,\tau\right)  -u^{2}\left(  x,t\right)  /2,\\
x=\sigma-\eta\left(  x,t\right)  , & t=\tau+u\left(  x,t\right)  ,
\end{array}
\]
gives us the solution $\left(  \eta\left(  x,t\right)  ,u\left(  x,t\right)
\right)  $ of (\ref{SWE1}) up to error $O\left(  \varepsilon\right)  $. This
solution remains valid as long as the non-breaking condition (\ref{nonbreaking for all x,t}) is satisfied. To obtain the solution for given values of $(x,t)$, Newton-Raphson iterations could be employed \citep{Synolakis87,KANOGLU04}. An example of data projection is depicted in figure \ref{fig:gammaproject}.

If the wave reaches a gradient catastrophe (i.e. it breaks) at some point then
our SWE (\ref{SWE}) is no longer valid and some other approximations of the
Navier-Stokes equations should be used \citep[e.g.][]{Johnson}.

\section{Example of power-shaped bays}\label{ubays}
In this section we apply the algorithm from Subsection \ref{algorithm} to the
case when $f\left(  y\right)  \sim\left\vert y\right\vert ^{m},\ 0<m\leq
\infty$ (called a power-shaped bay). We then have explicitly $c\left(
\sigma\right)  =\omega\sqrt{\sigma}$, where $\omega=\sqrt{m/(m+1)}$.

\subsection{Solution by data projection techniques}

In this subsection we consider the case of IC. The linear system
(\ref{standard IVP}) then reads
\begin{equation}
\left\{
\begin{array}
[c]{ccc}
\partial_{\tau}\psi+\omega^{2}\sigma\partial_{\sigma}\varphi+\varphi & = & 0\\
\partial_{\tau}\varphi+\partial_{\sigma}\psi & = & 0\\
\varphi\left(  \sigma,0\right)   & = & \varphi_{n}\left(  \sigma\right)  \\
\psi\left(  \sigma,0\right)   & = & \psi_{n}\left(  \sigma\right)  \\
\left\vert \varphi\left(  0,\tau\right)  \right\vert ,\left\vert \psi\left(
0,\tau\right)  \right\vert  & < & \infty\\
\varphi\left(  \infty,\tau\right)  ,\psi\left(  \infty,\tau\right)   & = & 0
\end{array}
\right.  ,\label{U bays}
\end{equation}
where we have merely supplemented the IVP with physically motivated BC.

Compute $\varphi_{n},$ $\psi_{n}$ in (\ref{U bays}) by (\ref{series}) with $n$
sufficiently large to provide a negligible error. The new equivalent problem
admits an explicit solution in terms of Bessel functions. One can merely do it
by the Hankel transform. Instead, we however use the explicit formulas readily
available from our \citet{Anderson}. For the reader's convenience we outline
the derivation from \citet{Anderson}. Reduce the system of PDEs in
(\ref{U bays}) to the single linear PDE
\begin{equation}
\partial_{\tau}^{2}\psi=\omega^{2}\sigma\ \partial_{\sigma}^{2}\psi
+\partial_{\sigma}\psi.\label{eq:psi_lin}
\end{equation}
Note that the differential operation on the right hand side of
(\ref{eq:psi_lin}) has a regular singular point at $\sigma=0$. This means that
(\ref{eq:psi_lin}) has a bounded and an unbounded solution at $\sigma=0$. The
latter one is discarded by the boundedness condition at $\sigma=0$. By the
standard Hankel transform techniques then for the general solution to
(\ref{eq:psi_lin}) we have
\begin{equation}
\psi(\sigma,\tau)=\sigma^{-\frac{1}{2m}}\int_{0}^{\infty}\left\{
a(k)\cos(\omega k\tau)+b(k)\sin(\omega k\tau)\right\}  J_{1/m}\left(
2k\sqrt{\sigma}\right)  dk,\label{eq:psi}
\end{equation}
where $J_{\nu}$ is the Bessel function of the first kind of order $\nu$ and
$a(k)$ and $b(k)$ are arbitrary functions determined by IC. It follows then
from (\ref{U bays}) and (\ref{eq:psi}) that
\begin{equation}
\varphi(\sigma,\tau)=\frac{1}{\omega}\sigma^{-\frac{1}{2m}-\frac{1}{2}}
\int_{0}^{\infty}\left\{  a(k)\sin(\omega k\tau)-b(k)\cos(\omega
k\tau)\right\}  J_{1/m+1}\left(  2k\sqrt{\sigma}\right)  dk.\label{eq:phi}
\end{equation}
We note that the apparent singularities at $\sigma=0$ in (\ref{eq:psi}) and
(\ref{eq:phi}) are actually removable due to asymptotic properties of the
Bessel function of the first kind around $0$.

The functions $a$ and $b$ can now be found from the IC by applying the inverse
Hankel transform to (\ref{eq:psi}) and (\ref{eq:phi}):
\begin{subequations}
\label{eq:a and b}
\begin{align}
a(k)  &  =2k\int_{0}^{\infty}\psi_{n}(s)s^{\frac{1}{2m}}J_{\frac{1}{m}}\left(
2k\sqrt{s}\right)  ds,\label{a}\\
b(k)  &  =-2\omega k\int_{0}^{\infty}\varphi_{n}(s)s^{\frac{1}{2m}+\frac{1}
{2}}J_{\frac{1}{m}+1}\left(  2k\sqrt{s}\right)  ds \label{b}
\end{align}
where $\psi_{n}(s),\varphi_{n}\left(  s\right)  $ are computed by
(\ref{series}). Thus $\psi$ and $\varphi$ are completely determined and
(\ref{U bays}) is explicitly solved.

In particular, for waves with zero initial velocity (i.e. $u_{0}=0$ and hence
$\varphi_{0}=0$) $b(k)=0$ and
\end{subequations}
\begin{equation}
a(k)=2k\int_{x_{0}}^{\infty}\left[  x+\eta_{0}(x)\right]  ^{\frac{1}{2m}
}J_{1/m}\left(  2k\sqrt{x+\eta_{0}(x)}\right)  \left[  1+\eta_{0}^{\prime
}(x)\right]  \eta_{0}(x)\ dx,\label{eq:a zero u}
\end{equation}
where we have used a simple change of variables to return back to the physical
space, and $x_{0}$ is the maximum run-up (i.e. $x_{0}+\eta_{0}(x_{0})=0$).

\subsection{\bigskip Finite power-shaped bay\label{finite}}
Here we consider a power-shaped bay $f\left(  y\right)  \sim\left\vert y\right\vert ^{m},\ 0<m\leq
\infty$ of finite length $l$ and set up some boundary conditions at $l$, e.g. see \citep{Harris15}
\[
\eta\left(  l,t\right)  =\eta_b\left(  t\right)  ,\ \ \ u\left(  l,t\right)
=u_b\left(  t\right)  .
\]
Since $B$ is a constant matrix and
\[
A^\prime = \frac{d}{d\sigma}\left(
\begin{array}
[c]{cc}
0 & 1\\
\omega^2\sigma & 0
\end{array}
\right)  =\left(
\begin{array}
[c]{cc}
0 & 0\\
\omega^2 & 0
\end{array}
\right)
\]
is also a constant matrix, the results of Subsection \ref{BVP general} apply. Recall that the constant $\omega=\sqrt{m/(m+1)}$. For the curve we have
\[
\Gamma_b=\left\{  \left(  \sigma_b\left(  \tau\right)  ,\tau\right)  |\gamma_b\left(  \tau\right)  \geq0\right\}  ,
\]
where $\gamma_b\left(  \tau\right)  $ is the inverse function of
$\tau=t-u_b\left(  t\right)  $ and
\[
\sigma_b\left(  \tau\right)  =l+\eta_b|_{\gamma_b\left(  \tau\right)  }.
\]
Eq. (\ref{k partial x of U}) reads
\[
\partial_{\sigma}^{k}\Phi|_{\Gamma_b}=D^{-1}\left(  \frac{d}{d\tau}
+B_{k-1}\right)  D^{-1}\left(  \frac{d}{d\tau}+B_{k-2}\right)  ...D^{-1}\left(
\frac{d}{d\tau}+B_{0}\right)  \Psi_{0},
\]
where
\[
D\equiv\left(
\begin{array}
[c]{cc}
\sigma_b^{\prime}\left(  \tau\right)   & -1\\
-\omega^2 \sigma_b\left(  \tau\right)   & \sigma_b^{\prime}\left(  \tau\right)
\end{array}
\right)  ,\ B_{k}=\left(
\begin{array}
[c]{cc}
0 & 0\\
1+k\omega^2 & 0
\end{array}
\right)  .
\]
Eq. (\ref{projection for BC}) for our case now yields\newline
\[
\widetilde{\Psi}_{0}(\tau)=\sum_{k=0}^{n}\frac{(\sigma_0-\sigma_b\left(\tau\right))^{k}}{k!}\partial_{\sigma}^{k}\Phi|_{\Gamma_b}.
\]
Thus the problem with a floating boundary condition is reduced to a fixed one
\begin{equation}
\left\{
\begin{array}
[c]{ccc}
\partial_{\tau}\psi+\omega^{2}\sigma\partial_{\sigma}\varphi+\varphi & = & 0\\
\partial_{\tau}\varphi+\partial_{\sigma}\psi & = & 0\\
\varphi\left(  \sigma,0\right)   & = & \varphi_{n}\left(  \sigma\right)  \\
\psi\left(  \sigma,0\right)   & = & \psi_{n}\left(  \sigma\right)  \\
\varphi\left(\sigma_0,\tau\right)  & = & \widetilde\varphi_{b}\left(  \tau\right)  \\
\psi\left(\sigma_0,\tau\right)   & = & \widetilde\psi_{b}\left(  \tau\right)
\end{array}
\right.\label{U bays2}
\end{equation}
Here, $\widetilde\varphi_b\left(  \tau\right), \widetilde\psi_b\left(\tau\right)$ are components of the vector $\widetilde{\Psi}_{0}(\tau)$. The value of $\sigma_0$ is chosen to be compatible with the IC, or $\sigma_0=l+\eta_0(l,0)$.

Note that an arbitrary boundary condition at $l$ need not produce a bounded solution to (\ref{U bays2}), i.e. we may have an infinite run-up (the energy will of course be finite). The physical relevance of such solutions is debatable but they can be avoided by imposing a compatibility condition for $\left(  \eta_b, u_b\right)  $. Such compatibility conditions are dictated by the underlying physics \citep[e.g.][]{Antuono07,Antuono10}.

Assuming that $\eta_b$ and $u_b$ are compatible, we can then handle
\begin{equation}
\left\{
\begin{array}
[c]{ccc}
\partial_{\tau}^{2}\psi&=&\omega^{2}\sigma\ \partial_{\sigma}^{2}\psi+\partial_{\sigma}\psi\\
\psi\left(  \sigma,0\right)   & = & \psi_{n}\left(  \sigma\right)  \\
\psi_\tau\left(\sigma,0\right)&=&-\omega^{2}\sigma\partial_{\sigma}\varphi_n(\sigma)-\varphi_n(\sigma)\\
\psi\left(\sigma_0,\tau\right)   & = & \widetilde\psi_{b}\left(  \tau\right)\\
|\psi\left(0,\tau\right)|  & < & \infty
\end{array}
\right.\label{U bays3}
\end{equation}
by a Bessel-Fourier expansion as follows.

By introducing the change of variables $\zeta^2=\sigma/\sigma_0$ and $\psi(\sigma,\tau) = \zeta^{-\gamma}\theta(\zeta,\tau)+\widetilde\psi_{b}(\tau)$, where $\gamma=1/m$, the wave equation is obtained
\begin{equation}
\partial^2_\tau\theta=k^2\left(\partial^2_\zeta\theta+\frac{1}{\zeta}\partial_\zeta\theta-\frac{\gamma^2}{\zeta^2}\theta\right)-\zeta^\gamma\widetilde\psi_{b}^{\prime\prime}(\tau),\label{HCBzeta}
\end{equation}
which admits a solution in terms of the Bessel functions $J_\gamma$ of order $\gamma$. Here, the prime denotes a derivative with respect to $\tau$, $k^2=\omega^2/4\sigma_0$, and $\zeta\in[0,1]$. The boundary condition at $\sigma=\sigma_0$ is transformed to $\theta(1,\tau)=0$. Next, the Fourier-Bessel decomposition is employed so that
\begin{equation}
\theta(\zeta,\tau)=\sum_{n=1}^{\infty}c_n(\tau)J_\gamma(j_n\zeta).\label{BF}
\end{equation}
To solve for the coefficients, we substitute (\ref{BF}) into (\ref{HCBzeta}) and use an orthogonality property of  Bessel functions to obtain a set of ordinary differential equations for each coefficient $c_n$:
\begin{equation}
c^{\prime\prime}_n(\tau)+(j_nk)^2c_n(\tau)=-\frac2{J_{\gamma+1}^2(j_n)}\widetilde\psi_{b}^{\prime\prime}(\tau)\int_0^1\zeta^{1+\gamma}J_\gamma(j_n\zeta)d\zeta.\label{ccoeff}
\end{equation}
The initial conditions in (\ref{U bays3}) could be cast to yield the initial conditions for $c_n$ such that
\[
\begin{array}
[c]{ccc}
c_n(0)&=&\frac{2}{J_{\gamma+1}^2(j_n)}\int_0^1{\zeta}^{1+\gamma}J_\gamma(j_n\zeta)\left[\psi_n(\sigma_0\zeta^2)-\widetilde\psi_b(0)\right]d\zeta,\\
c_n^\prime(0)&=&-\frac{2}{J_{\gamma+1}^2(j_n)}\int_0^1{\zeta}^{1+\gamma}J_\gamma(j_n\zeta)\left[\frac12\omega^2\zeta\partial_\zeta\varphi_n(\sigma_0\zeta^2)+\varphi_n(\sigma_0\zeta^2)+\widetilde\psi_b^\prime(0)\right]d\zeta.
\end{array}
\]
Finally, we express $\psi$ and $\phi$ in terms of variables $(\sigma, \tau)$ as
\begin{eqnarray}
\psi(\sigma,\tau)&=&\left(\frac{\sigma_0}{\sigma}\right)^{\frac{1}{2m}}\sum_{n=1}^{\infty}c_n(\tau)J_{\frac{1}{m}}(j_n\sqrt{\sigma/\sigma_0})+\widetilde\psi_b(\tau),\label{BesFor1}\\
\varphi(\sigma,\tau)&=&\frac{1}{2\sigma_0}\left(\frac{\sigma_0}{\sigma}\right)^{\frac{1}{2m}+\frac{1}{2}}\sum_{n=1}^{\infty}j_nd_n(\tau)J_{\frac{1}{m}+1}(j_n\sqrt{\sigma/\sigma_0}),\label{BesFor2}
\end{eqnarray}
where,
\[
d_n(\tau) = \int_0^\tau{c_n(\lambda)d\lambda}.
\]
We calculate runup and rundown of the Gaussian wave in a V-shaped bay ($\omega=1/\sqrt2$) using equations (\ref{BesFor1})-(\ref{BesFor2}) in Subsection \ref{finite}.

\subsection{Comparison to previous results}

As we have mentioned in the introduction, the problem of adjusting the CG
transform techniques to an arbitrary nonzero initial velocity has been
approached by many authors. We will not discuss the complete history of the
problem and by the same token will not give an attempt to review the extensive
literature. Instead, we concentrate only on the most important contributions
where the interested reader can find further references.

The first significant result to this effect appeared in \citet{Yeh03}. It was
then improved in \citet{KANOGLU04} and \citet{Kanoglu06}, where the Green's
function approached was employed. More specifically, for the plane beach
($m=\infty$) under the assumption that $\sigma=x$ a certain solution formula was
derived. It can be shown (see our \citep{Nicolsky18} for the details) that
this solution is exact only if $\varphi_{0}^{\prime}\left(  \sigma\right)  $
$=0$. However, for near shore waves with large initial velocities such
solution may produce some artifacts. If $\varphi_{0}^{\prime}\left(
\sigma\right)  =0$ then $D=I$ and (\ref{series}) simplifies to read
\begin{align}
\left(
\begin{array}
[c]{c}
\varphi_{n}\left(  \sigma\right)  \\
\psi_{n}\left(  \sigma\right)
\end{array}
\right)   &  =\sum_{k=0}^{n}\frac{\varphi_{0}^{k}\left(  \sigma\right)  }
{k!}\Delta^{k}\left(
\begin{array}
[c]{c}
\varphi_{0}\left(  \sigma\right)  \\
\eta_{0}|_{\gamma\left(  \sigma\right)  }+\varphi_{0}^{2}\left(
\sigma\right)  /2
\end{array}
\right)  \nonumber\\
&  =\left(
\begin{array}
[c]{c}
\varphi_{0}\left(  \sigma\right)  \\
\eta_{0}|_{\gamma\left(  \sigma\right)  }+\varphi_{0}^{2}\left(
\sigma\right)  /2
\end{array}
\right)  -\varphi_{0}\left(  \sigma\right)  \left(
\begin{array}
[c]{c}
0\\
\left(  \eta_{0}|_{\gamma\left(  \sigma\right)  }\right)  ^{\prime}
+\varphi_{0}\left(  \sigma\right)
\end{array}
\right)  \label{siml series}\\
&  +\dfrac{\varphi_{0}^{2}\left(  \sigma\right)  }{2}\left(
\begin{array}
[c]{c}
0\\
\left(  1+c^{2}\left(  \sigma\right)  \right)  \left(  \eta_{0}|_{\gamma
\left(  \sigma\right)  }\right)  ^{\prime}+c^{2}\left(  \sigma\right)  \left(
\eta_{0}|_{\gamma\left(  \sigma\right)  }\right)  ^{\prime\prime}
\end{array}
\right)  +...\nonumber
\end{align}
It is a straightforward (but quite involved) exercise to show that combining
(\ref{eq:psi})-(\ref{b}), and (\ref{siml series}) yields $\psi\left(
\sigma,\tau\right)  $ which considers with the solution in \citet{Kanoglu06}
up to $O\left(  \varphi_{0}^{\prime}\left(  \sigma\right)  \right)  $.
However, as numerical simulations in the next subsection show, our scheme runs
smoothly without the assumption that $\varphi_{0}^{\prime}\left(
\sigma\right)  $ \ is small. Incidentally, (\ref{siml series}) demonstrates
the analytical complexity of our data projection method.

In very interesting papers \citep{Antuono07,Antuono10} perturbation techniques
are used to deal with boundary value problems. However, such techniques could
also be adjusted to the IVP but would require certain smallness of the BC and IC.

The IVP has also been considered in the context of parabolic bays where $m=2$.
\citet{Didem} derived an exact traveling wave solution of the IVP in parabolic
bays for waves with zero initial velocity. In parabolic bays where $m=2$,
using the identity $J_{1/2}(x)=\sqrt{2/(\pi x)}\sin(x)$ along with other
trigonometric identities, (\ref{eq:psi}) and (\ref{eq:a zero u}) reduce (again
after quite involved computations) to
\begin{align*}
&  \psi(\sigma,\tau)\\
&  =\frac{1}{2\sqrt{\sigma}}\left\{  \Theta\left(  \sqrt{\sigma}+\frac{\tau
}{\sqrt{6}}\right)  +\Theta\left(  \sqrt{\sigma}-\frac{\tau}{\sqrt{6}}\right)
\theta(\sqrt{\sigma}-\frac{\tau}{\sqrt{6}})-\Theta\left(  \frac{\tau}{\sqrt
{6}}-\sqrt{\sigma}\right)  \theta(\frac{\tau}{\sqrt{6}}-\sqrt{\sigma
})\right\}
\end{align*}
where $\Theta(\zeta)=\zeta\eta_{0}(\gamma(\zeta)),$ $\theta$ is the Heaviside
function, and $\gamma(\sigma)$, as before, is given implicitly by $x+\eta
_{0}(x)=\sigma$. This solution is identical to the one given by \citet{Didem}
under the change of variables $\sigma=\sqrt{6s}$, $\lambda=-\tau$,
$\varphi=\phi_{\sigma}/\sigma$ and $\psi=\phi_{\lambda}/3$.

IBVP in the same context have also been treated by many authors, see \citep[e.g.][]{Synolakis87,KANOGLU04,Anderson}
and  the literature cited therein, where floating points are fixed by assuming a certain negligible difference between $\sigma$ and $x$ far away from the shore. Our approach does not require such assumptions.

\section{Numerical Verification of the data projection method}\label{Numerics}

\subsection{Verification for the initial value problem}

In this subsection, we numerically verify our data projection method for the
initial value problem (\ref{U bays}) by considering runup of the Gaussian wave
\begin{equation}
\eta_{0}(x)=ae^{-b(x-x_{0})^{2}}, \label{eq:gauss}
\end{equation}
in a bay of the parabolic shape ($m=2$). To do that we consider initial
condition $\left(  \eta_{0},0\right)  $ (i.e. with zero initial velocity) and
run it by the standard CG to the maximum runup at $t=t_{r}$. While modeling
the runup $\eta(x,t_{r})$, we record $\left(  \eta\left( x,t_{*}\right) ,
u\left( x,t_{*}\right)  \right)  $ at some time $t_{*}<t_{r}$. We then set up
a new IVP\ with IC $\left(  \eta\left( x,t_{*}\right) , u\left( x,t_{*}
\right)  \right)  $ and run it by our method. Both solutions (via the standard
CG and the new IVP) are expected to show an excellent agreement for $t_{r}{\ge}t>t_{*}
$. Results of the comparison are provided below.

\begin{figure}[th]
\center
\includegraphics[width=0.9\textwidth]{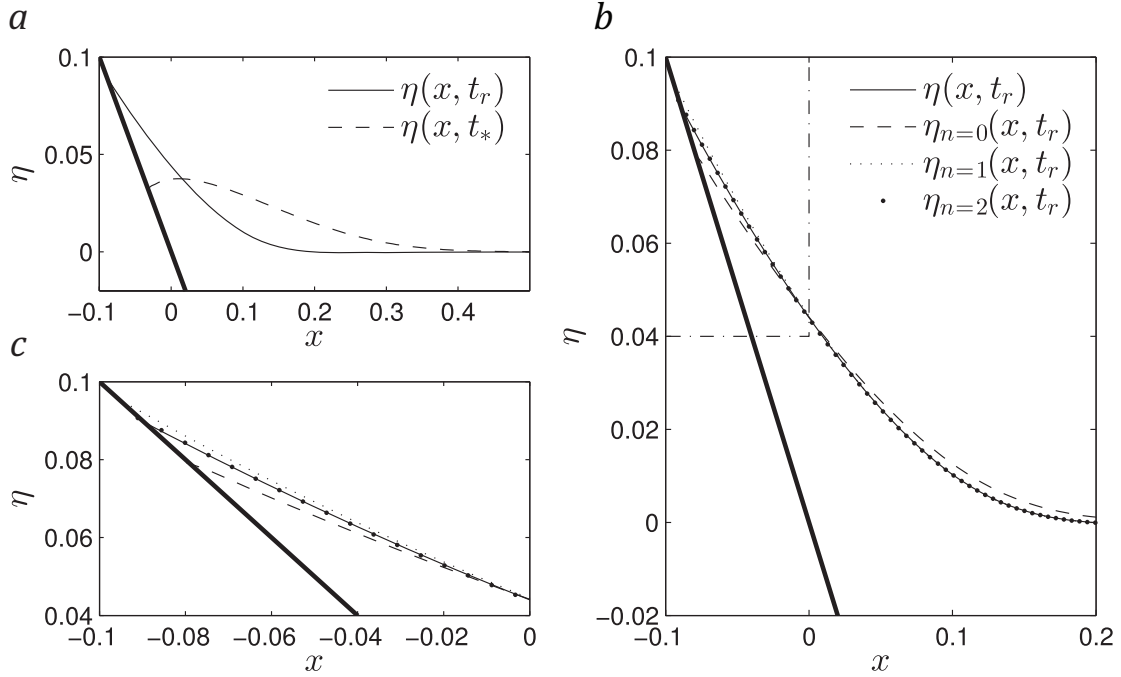}\caption{a: profiles of the water level $\eta$ for the initial condition: a zero-velocity Gaussian wave given by (\ref{eq:gauss}) with $a=0.017$, $b=4.0$ and $x_{0}=1.69$ running up a parabolic bay ($m=2$). Profile $\eta(x,t_{*})$ is used in the proposed data projection method to solve a non-zero initial velocity problem. b: Comparison of the water level at $t=t_{r}$ for various approximations of $\Phi_{n}$. c: Zoomed-in comparison of water level near the shoreline, i.e. within the dashed rectangle is shown in plot b.}\label{fig:validation1}
\end{figure}

In particular, as in \citep{Kanoglu06,Nicolsky18} we consider an initial Gaussian
wave with $a=0.017$, $b=4.0$ and centered at the distance of $x_{0}=1.69$ from
the shore. In this case, the maximum runup occurs at $t_{r}\approx2.908$, and
we choose $t_{*}=2$, when the wave is approximately half the way to its
maximum runup on the shore (and where of course $u(x,t_{*})\ne0$). Figure
\ref{fig:validation1}a displays wave profile at the time of maximum runup
$t_{r}$ and at the moment $t_{*}$. We launch our method forming the projected
IC by (\ref{series}) with various degrees of approximation $n=0,1,2$ and apply
formulas (\ref{eq:psi}-\ref{eq:phi}) to model the wave propagation until
$t=t_{r}$. We note that $c^{2}(\sigma)=2/3\sigma$. To compute projections of the IC and BC, we use recursive formulae and compute the first order derivatives by the finite differences of the second order accuracy wherever possible.

Comparison of the water level profiles $\eta(x,t_{r})$ and $\{\eta
_{n=k}(x,t_{r})\}_{k=0}^{2}$ at the moment of maximum runup is shown in Figure
\ref{fig:validation1}b. Unlike to results by \citet{Nicolsky18}, where the
zeroth approximation, $n=0$, was adequate to capture the wave profile at
$t=t_{r}$, here the zeroth approximation shows a visible deviation from
$\eta(x,t_{r})$ due to a larger initial velocity $\eta(x,t_{*})$ used in the
data projection method. Notice that for the high-order approximations,
$n=1,2$, the match between the water profiles at $t=t_{r}$ improves and
becomes satisfactory. The convergence of approximations, $\{\eta_{n=k}
(x,t_{r})\}_{k=0}^{2}$, is demonstrated near the tip of the wave; an area
within the dashed rectangle in Figure \ref{fig:validation1}b is shown in
Figure \ref{fig:validation1}c. One may notice that the zeroth approximation
$\eta_{n=0}(x,t_{r})$ undershoots the runup, the first-order approximation $\eta_{n=1}(x,t_{r})$ overshoots
and the second order $\eta_{n=2}(x,t_{r})$ almost overlaps $\eta(x,t_{r})$. Other higher order
approximations (not shown for the sake of clarity) provide a nearly exact
match to $\eta(x,t_{r})$. This demonstrates an efficacy of the proposed method
to project the solution forward from the given initial conditions.

\subsection{Verification for the boundary value problem\label{Verif BVP}}

In this subsection, we numerically verify our data projection method for the
boundary value problem (\ref{U bays3}) by considering runup of the Gaussian wave
(\ref{eq:gauss}) in a V-shaped bay ($m=1$, $c^{2}(\sigma)=\left(  1/2\right)
\sigma$). Similar to the previous numerical experiment, we consider a
zero-velocity initial condition $\left(  \eta_{0},0\right)  $ and run it by
the standard CG to compute the maximum runup ($t=t_{r}$), rundown ($t=t_{d}$)
and the secondary runup $t=t_{s}$. The secondary runup of a Gaussian wave is a
feature of the V-shaped bay as it was noted by \citet{MHarris16,Nicolsky18}.
Now, while modeling the wave dynamics, we save the time history of $\eta\left(x,t\right)$ and $u\left(x,t\right)$ at some point $x=x_{\ast}$ near the shore (e.g. $x_{\ast}=0.15$) for $0\leq t\leq
t_{s}$. Note that in the previous subsection we recorded the snapshot of the
wave dynamics to setup the IVP. Here, we use the saved time history $\left(
\eta\left(  x_{\ast},t\right)  ,u\left(  x_{\ast},t\right)  \right)  $ to set
up a new BVP and run it by our method (assuming in the data projection algorithm that $l=x_{\ast}$, $\eta_b(t)=\eta\left(x_{\ast},t\right)$ and $u_b(t)=u\left(x_{\ast},t\right)$). Both solutions via the standard CG and the new BVP are again to show an excellent agreement for $t<t_{s}$.

\begin{figure}[ptb]
\center
\includegraphics[width=0.9\textwidth]{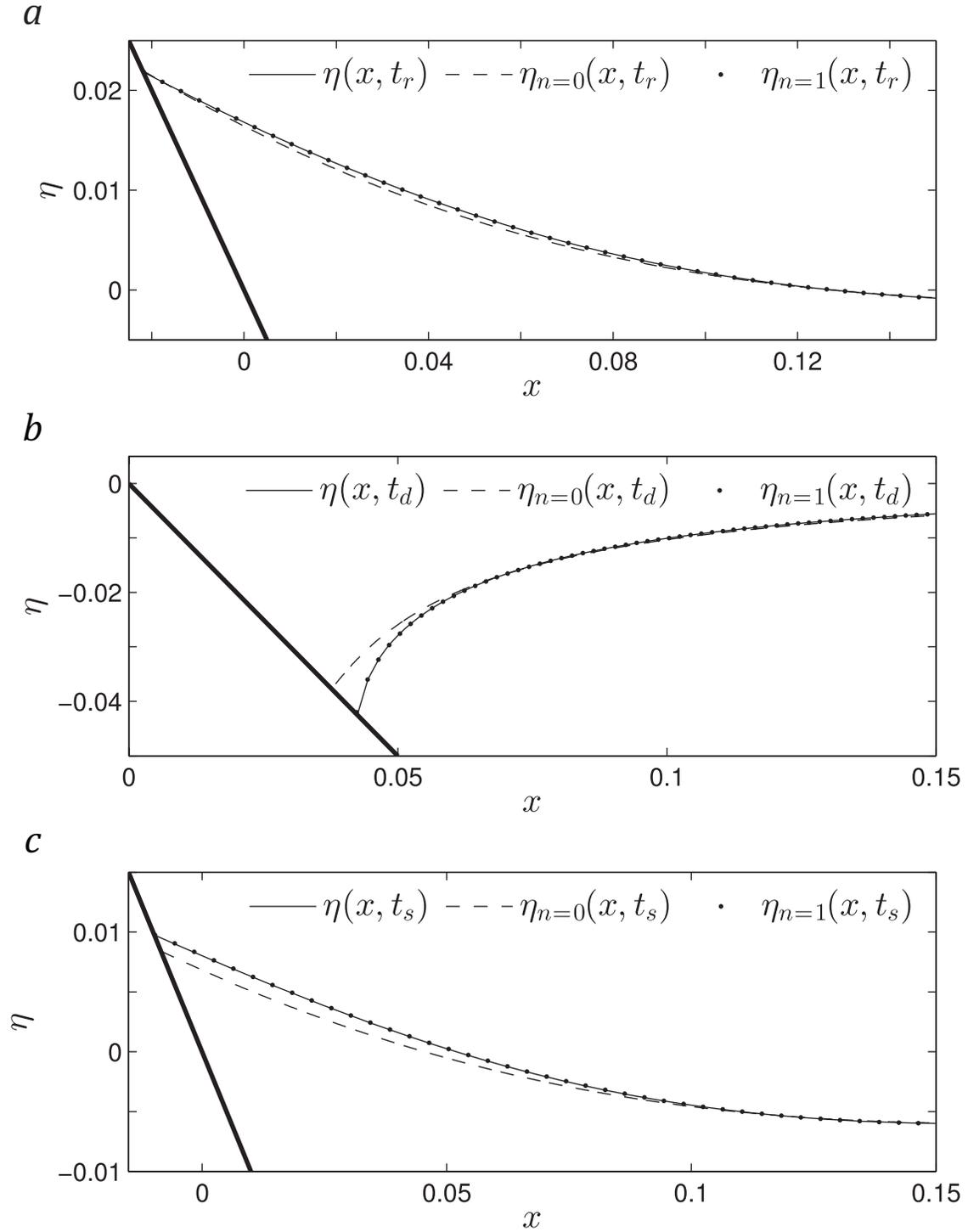}\caption{Comparison of the water level profiles $\eta(x,t)$ and $\{\eta_{n=k}(x,t)\}_{k=0}^{2}$ at the moment of (a) maximum runup $t=t_{r}$, (b) rundown $t=t_{d}$ and (c) the secondary runup $t=t_{s}$ in the V-shaped bay ($m=1$).}\label{fig:validation2}
\end{figure}

Similar to the previous experiment, we consider an initial Gaussian wave with
the same characteristics, but the amplitude is reduced ($a=0.017$) to have a highest
non-breaking wave throughout the simulation. Comparisons between the two
solutions at the moments of maximum runup, rundown and the secondary runup are
provided in Figure \ref{fig:validation2}. The water level $\eta_{n=0}$ for the
zeroth approximation shows a discrepancy with $\eta$. However, results for the
next order of approximation, i.e. $n=1$, match the true solution almost
exactly at the two run-ups and rundown. Other higher orders rapidly converge
and provide nearly exact match.

\subsection{Modeling shore dynamics for the incident N-wave}
\begin{figure}[ptb]
\center
\includegraphics[width=0.9\textwidth]{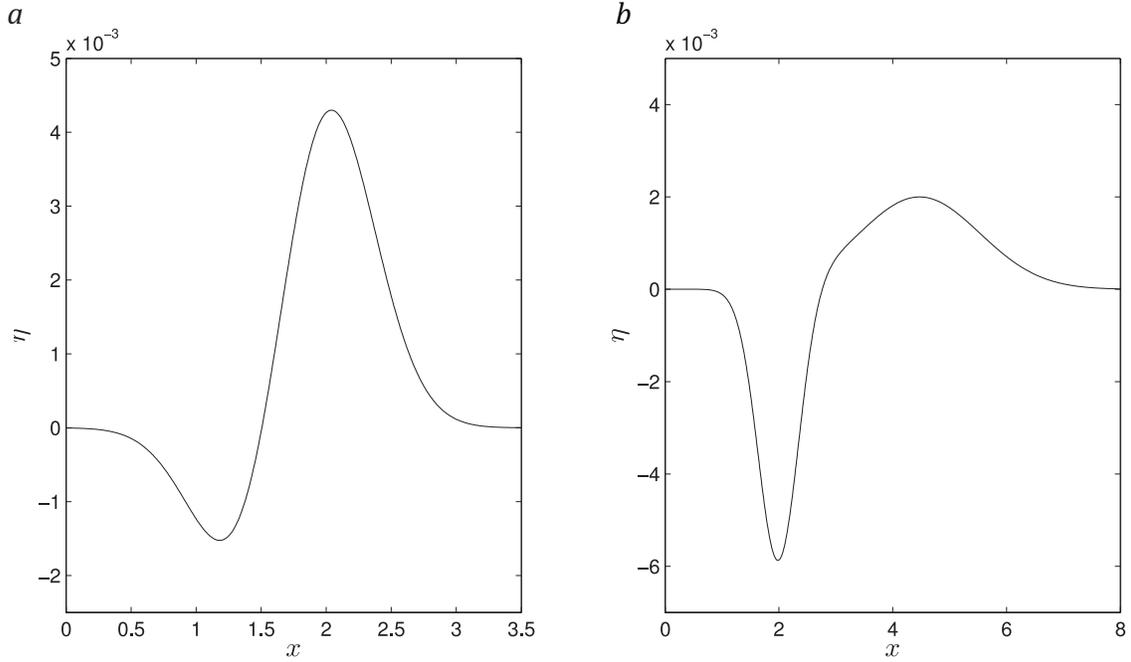}\caption{Initial leading-depression N-waves with the geometries similar to \citet{Yeh03,KANOGLU04}: (a) $a_{1}=0.005$, $b_{1}=3.5$, $x_{1}=1.9625$, $a_{2}=-0.0025$, $b_{2}=3.5$, and $x_{2}=1.4$. (b) $a_{1}=0.002$, $b_{1}=0.4444$, $x_{1}=4.4709$, $a_{2}=-0.006$, $b_{2}=4.0$, and $x_{2}=1.9884$}\label{fig:nwaves}
\end{figure}

\begin{figure}[ptb]
\center
\includegraphics[width=0.9\textwidth]{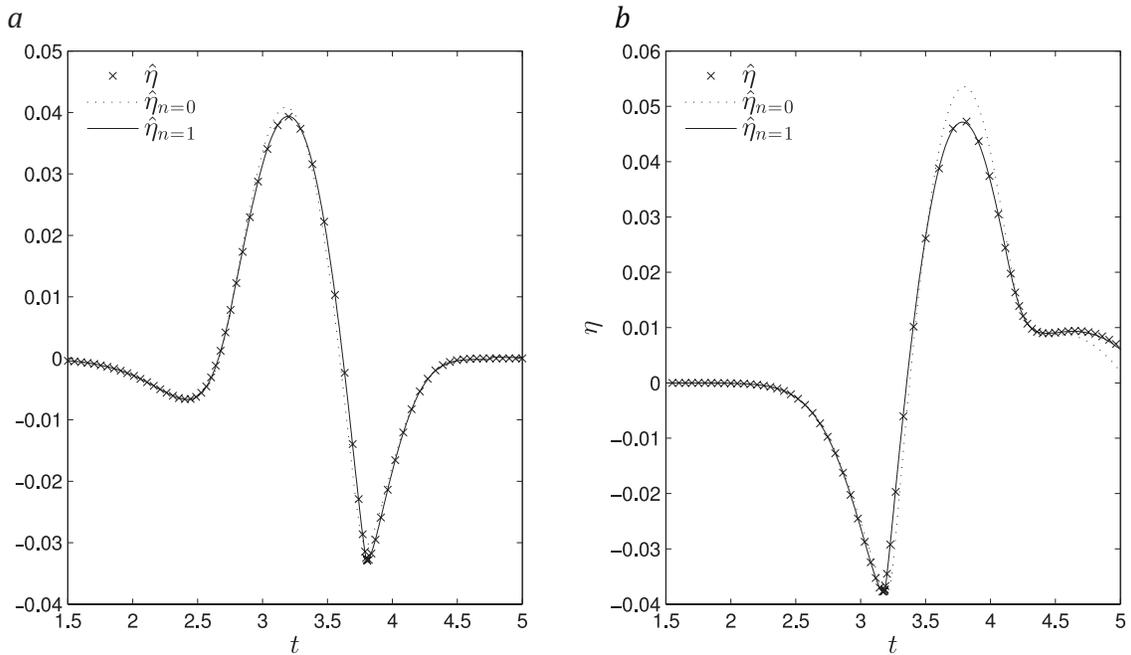}\caption{Comparison of the shoreline dynamics $\hat\eta$ for the leading-depression N-waves in the parabolic bay ($m=2$). Dynamics in (a) and (b) correspond to cases in Figure \ref{fig:nwaves}.}\label{fig:shoreline_nwaves}
\end{figure}

To illustrate efficacy of the proposed method for BVP (\ref{U bays3}) we also consider runup of N-shaped
waves in the parabolic bay ($m=2$)
\begin{equation}
\eta_{0}(x)=a_{1}e^{-b_{1}(x-x_{1})^{2}}+a_{2}e^{-b_{2}(x-x_{2})^{2}}.
\label{eq:nwave}
\end{equation}
In particular, we consider two leading-depression N-shaped waves with the
geometries similar to those in \citep{Yeh03}. Both waves have zero initial
velocities and their profiles are shown in Figure \ref{fig:nwaves}. As in the
previous subsection, we model the wave dynamics using the standard CG and
record the water level as well as the velocity at some point $x=x_{*}$ near the shore, e.g. at
$x_{*}=0.15$. The recorded history of $\left( \eta\left( x_{*},t\right) ,
u\left( x_{*},t\right) \right) $ is then again used to set up a new BVP with $l=x_{*}$, $\eta_b(t)=\eta\left( x_{*},t\right)$ and $u_b(t)=u\left( x_{*},t\right)$ to model shoreline dynamics. Figure \ref{fig:shoreline_nwaves} shows comparison between
the shoreline $\hat\eta$ computed with the standard CG and those $\{\hat
\eta_{n=k}\}$ obtained from different order approximations, i.e. $k=0,1$.
Notice that even for the zeroth approximation, $n = 0$, the match between the
shorelines is rather good. However, some discrepancy exists at the maximum
runup, the zeroth order overestimates the maximum runup. However, the first
order approximation, $n=1$, nearly exactly matches the true solution during
the runup and rundown.

\subsection{Contribution of the wave velocity to runup}
\begin{figure}[th]
\includegraphics[width=0.9\textwidth]{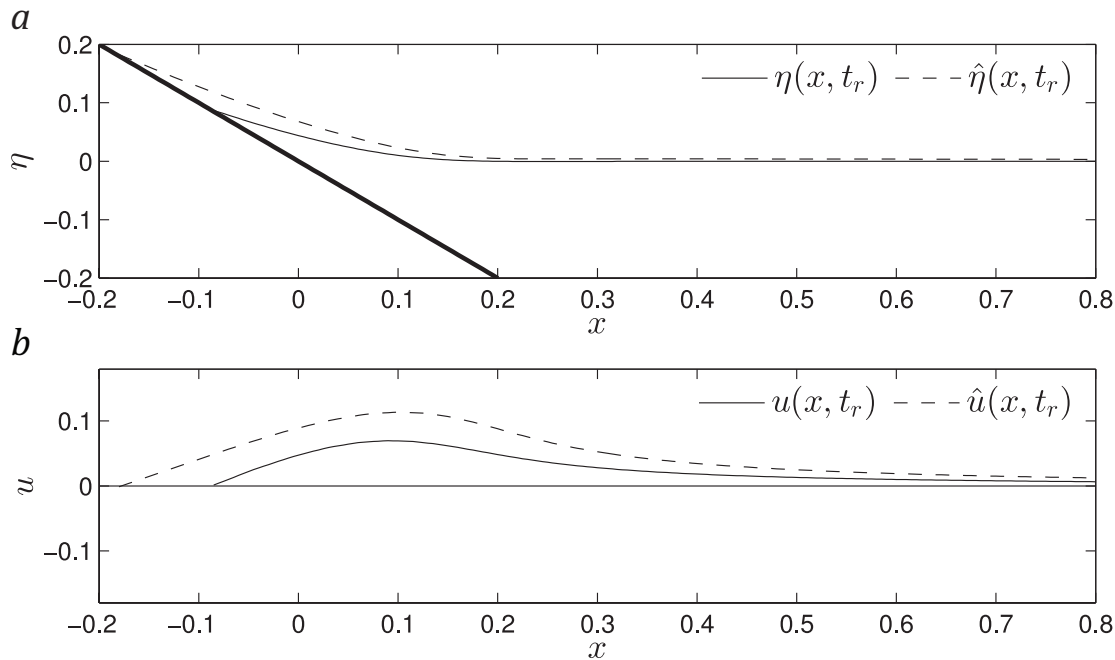} \caption{Comparison of the water level (a) and water velocity (b) at the time $t=t_{r}$ of maximum runup. Quantities marked with the symbol $\hat{}$ are computed in the case when the initial velocity $u_{0}(x)=-2\sqrt{(m+1)/m}\left( \sqrt{\eta+h}-\sqrt{h}\right)$.}\label{fig:Zero_vs_NZero}
\end{figure}

We conclude this section with illustrating a physical effect showing how the
runup increases when initial velocity is present. When initial velocity is
absent the initial wave splits and propagates in both directions from the
source region, i.e. towards shore and away from it. It was shown by
\citet{Didenkulova11b} that in a flat bottom fjords with the power-shaped
cross-section, the wave propagates towards the shore when the initial velocity
satisfies
\begin{equation}
u_{0}(x)=-2\sqrt{(m+1)/m}\left(  \sqrt{\eta+h}-\sqrt{h}\right)  .
\label{eq:nonzerovelocity}
\end{equation}
We use this approximation in the following numerical experiment. As before, we consider the parabolic bay ($m=2$) and take the same Gaussian wave with $u_{0}=0$ and with $u_{0}$ given by (\ref{eq:nonzerovelocity}) with $h=x$. In the former case,
the runup occurs $t_{r}\approx2.908$, whereas in the latter one the runup
happens $t_{r}\approx2.928$. That is, the runup occurs almost at the same
time, however, as one can see it in Figure \ref{fig:Zero_vs_NZero}, the
maximum run-up is almost twice as large for the non-zero initial velocity.
This result shows that long waves can be greatly amplified in heads of narrow
bays if the initial velocity nonzero.

\section{Conclusions}

Our method of data projection completely solves the problem of the effective
linearization of the shallow water equation (SWE)\ for any inclined bay
with IC (in arbitrary shaped bays) and BC (only in power-shaped bays) by means of the Carrier-Greenspan (CG) transform. Basing upon Taylor's formula in ``reverse'', given IC $\eta\left(  x,0\right)
=\eta_{0}\left(  x\right)  ,u\left(  x,0\right)  =u_{0}\left(  x\right)  $
(with $u_{0}\not =0$) for the (\ref{SWE1}), we find an equivalent IC
$\varphi\left(  \sigma,0\right)  =\varphi_{n}\left(  \sigma\right)  ,$
$\psi\left(  \sigma,0\right)  =\psi_{n}\left(  \sigma\right)  $ for the linear
SWE (\ref{SWE linearized S}) in the transformed space $\left(  \sigma
,\tau\right)  $ (hodograph plane). The initial value problem (IVP)
(\ref{standard IVP}) can then be easily solved analytically or numerically.
Performing the inverse CG transform solves the original IVP for the SWE
(\ref{SWE1}) to any order of accuracy. As is well-known, the main benefits of
(in fact, any) linearization are nearly instantaneous computations and
explicit analysis uncovering subtle properties of the system under
consideration. This method works songlessly for BC as well and hence for IC/BC
combined. The BC\ case though requires more attention than we were able to pay
in this paper. In particular, our approach may potentially be very useful in
the study of more complicated than inclined bathymetries treated in \citep{Synolakis87,Synolakis91}.

Our method, which becomes explicit for U-shaped bays, has potential
applications in tsunami wave modeling. Tsunami forecast models are extensively
verified against the analytical solutions of the SWEs \citep{Synolakis08},
primarily for the case of a plane beach. This solution allows further
analytical verification of tsunami models, with the extension of the solution
to 2-D bathymetries, allowing verification of tsunami models in realistic
settings. As local near-shore bathymetry significantly effects the run-up of
tsunami waves, and narrow bays can greatly amplify tsunami waves, the
verification of tsunami models in narrow bays is critical for protecting
coastal communities and infrastructure. Furthermore, 1-D nonlinear shallow
water theory has had significant developments in the past few years,
specifically in the context of narrow bays. In the realistic setting of
Alaskan fjords, 1-D theory has had similar runup predictions to full 2-D
tsunami models with significantly less computation time
\citep{Harris15a,Anderson}. 1-D theory can even present valid predictions in
splitting bays \citep{Raz2017}. With such progress, it is possible for 1-D
shallow water theory to be incorporated into global 2-D tsunami inundation
models, specifically in narrow bays and fjords. This will reduce computation
and forecasting time, potentially saving lives and resources.

Treating initial and boundary conditions for the SWE by means of the CG
transform opens new avenues in the analysis of much more realistic runup problems
for tsunami waves. In particular, we hope to develop a method of stitching
together different shallow water approximations describing different stages of
the tsunami wave propagation.

\vspace{10mm}

\textbf{Acknowledgments:} We would like to thank anonymous referees for careful reading of the manuscript and valuable comments, which have been very helpful in improving the manuscript. Also, we are grateful to Dillon Gillespie for his help with computations of the Bessel-Fourier expansion. Alexei Rybkin acknowledges support from National Science Foundation Grant (NSF)\ award DMS-1411560 and DMS-1716975. Dmitry Nicolsky acknowledges support from the Geophysical Institute, University of Alaska Fairbanks. Efim Pelinovsky acknowledges support by Laboratory of Dynamical Systems and Applications NRU HSE, by the Ministry of science and higher education of the RF grant ag. 075-15-2019-1931 and by FRBR grant 18-05-80019 and 20-05-00162. Maxwell Buckel was supported by the National Science Foundation Research Experience for Undergraduate program (Grant DMS-1411560).

\bibliographystyle{apalike}

\end{document}